\documentclass[prd,floatfix,nofootinbib]{revtex4}
\usepackage{epsfig}
\newcommand{\eq}[1]{Eq.~(\ref{#1})}

\newcommand{\fig}[1]{Fig.~{\ref{#1}}}

\newcommand{\be}{\begin{equation}}
\newcommand{\ee}{\end{equation}}
\newcommand{\bea}{\begin{eqnarray}}
\newcommand{\eea}{\end{eqnarray}}
\newcommand{\ben}{\begin{eqnarray*}}
\newcommand{\een}{\end{eqnarray*}}

\newcommand{\DS}{Schwinger--Dyson }
\newcommand{\BS}{Bethe--Salpeter }

\newcommand{\w}{\omega}
\newcommand{\e}{\varepsilon}
\newcommand{\al}{\alpha}
\newcommand{\ba}{\beta}
\newcommand{\ga}{\gamma}
\newcommand{\G}{\Gamma}
\newcommand{\de}{\delta}
\newcommand{\De}{\Delta}
\newcommand{\si}{\sigma}
\newcommand{\Si}{\Sigma}
\newcommand{\ro}{\rho}
\newcommand{\la}{\lambda}
\newcommand{\La}{\Lambda}

\newcommand{\pslash}{p \hskip -0.5em /}

\newcommand{\s}{\!\cdot\!}
\newcommand{\ov}[1]{\overline{#1}}
\newcommand{\dk}[1]{\,\,\,\raisebox{-0.4ex}{\large $\bar{}$}\!\!d\,{#1}\,}
\def\slr#1{\setbox0=\hbox{$#1$}           
   \dimen0=\wd0                                 
   \setbox1=\hbox{/} \dimen1=\wd1               
   \ifdim\dimen0>\dimen1                        
      \rlap{\hbox to \dimen0{\hfil/\hfil}}      
      #1                                        
   \else                                        
      \rlap{\hbox to \dimen1{\hfil$#1$\hfil}}   
      /                                         
   \fi}

\begin{document}

\title{{\scriptsize \tt IPPP/05/60~~~DCPT/05/118}\\
\vspace{5mm}
Probing Unquenching Effects in the Gluon Polarisation in Light Mesons}
\author{C.~S.~Fischer}
\email{christian.fischer@durham.ac.uk}
\affiliation{Institute for Particle Physics Phenomenology, University of Durham, South Road, Durham, DH1 3LE, UK}
\author{P.~Watson}
\email{peter.watson@theo.physik.uni-giessen.de}
\affiliation{Institute for Theoretical Physics, University of Giessen, Heinrich-Buff-Ring 16, 35392 Giessen, Germany}
\author{W.~Cassing}
\affiliation{Institute for Theoretical Physics, University of Giessen, Heinrich-Buff-Ring 16, 35392 Giessen, Germany}
\date{\today}

\begin{abstract}
We introduce an extension to the ladder truncated \BS equation for mesons and the rainbow truncated quark \DS equations 
which includes quark-loop corrections to the gluon propagator. This truncation scheme obeys the axialvector Ward-Takahashi 
identity relating the quark self-energy and the \BS kernel. Two different approximations to the Yang-Mills sector are used 
as input: the first is a sophisticated truncation of the full Yang-Mills \DS equations, the second is a phenomenologically 
motivated form.  We find that the spectra and decay constants of pseudoscalar and vector mesons are overall described well 
for either approach. Meson mass results for charge eigenstate vector and pseudoscalar meson masses are compared to lattice 
data. The effects of unquenching the system are small but not negligible. 
\end{abstract}

\maketitle

\section{Introduction}\label{sec:intro}

One of the longstanding problems of the theory of the strong interaction, Quantum Chromodynamics [QCD], is the 
description of hadronic bound states and resonances from first principles. Ironically it is perhaps the most 
interesting aspect of QCD, the phenomenon of confinement, that greatly hinders the development of our understanding 
of the experimental hadron spectrum: the individual components of these hadrons, dressed quarks and gluons, do 
not appear as physically observable states in nature. Thus, in order to study the internal structure of hadrons
one has to resort to theoretical frameworks that connect the fundamental fields of the theory with phenomenology
while preserving the basic low energy properties of QCD: confinement and dynamical chiral symmetry breaking.

Lattice simulations are not entirely satisfactory in this respect. They provide values for the global
properties of hadrons (masses, decay widths etc.,) which, when compared with experiment, reassure us that
QCD is indeed the theory of strong interactions. But they may not be capable to provide enough information 
on the internal structure of these hadrons, which is vital for our understanding of the dynamical aspects 
of low energy QCD.
An alternative field theoretical and relativistic method which is well suited to provide this information is 
the \DS and Bethe-Salpeter formalism (see, for example, \cite{iz,Alkofer:2000wg,Roberts:2000aa,Maris:2003vk} 
for a contemporary introduction to the topic). In principle this approach allows one to derive meson properties 
directly from the fundamental building blocks of the field theory, the Green's functions.

From a technical point of view, lattice simulations and the Green's functions approach are complementary to each 
other in several respects.
Lattice calculations are {\it ab initio} and thus contain all effects from quantum fluctuations. However, they are 
limited to a comparably small momentum range and suffer from finite volume effects in the infrared. Furthermore the 
implementation of realistic quark masses turns out to be extremely involved from the viewpoint of computational times 
and is probably not feasible in the near 
future. \DS equations [SDEs] can be solved analytically in the infrared and provide numerical solutions for a large momentum 
range. Together with the corresponding \BS framework all aspects of chiral symmetry are respected such that the properties 
and effects of light quarks can be determined systematically with reasonable effort. On the other hand, in order to obtain 
closed systems of equations one has to introduce suitable approximations of higher order n-point functions. These 
truncations have to be controlled by comparing results from the SDE-approach to corresponding lattice data. In turn, the
\DS approach provides important information in momentum and quark mass regions that are not (yet) accessible  
on the lattice.

One of the most important recent developments in the \BS formalism is the realisation that the chiral symmetry breaking 
pattern of the pseudoscalar meson sector is governed by the relationship between the quark propagator SDE and 
the \BS kernel supplied by the axialvector Ward-Takahashi identity \cite{Maris:1997hd}.  This observation has led to 
the development of phenomenologically motivated models to describe the light pseudoscalar and vector meson 
properties \cite{Maris:1999nt,Alkofer:2002bp}.  These models are constructed using the simplest truncation scheme 
(the rainbow/ladder truncation) known to satisfy the axialvector Ward-Takahashi identity. The Yang-Mills sector 
of QCD is replaced by an ansatz for the effective interaction between the quarks which is sufficiently strong 
to induce spontaneous chiral symmetry breaking and associated dynamical quark mass generation. This provided a robust 
starting point to determine various aspects of the light pseudoscalar and vector mesons like charge form 
factors \cite{Maris:2000sk,Volmer:2000ek}, electroweak and strong decays \cite{Maris:2001am,Jarecke:2002xd} and 
electroweak transitions \cite{Maris:2002mz,Ji:2001pj}. Exploratory studies go beyond the rainbow-ladder truncation 
by including corrections to the quark-gluon vertex \cite{Bender:2002as,Bhagwat:2004hn,Watson:2004kd}. 

Despite their phenomenological success, these models are unsatisfactory in an important respect. It is not clear, 
to what extent the structure of the Yang-Mills sector of QCD is represented by the ans\"atze used for the 
effective quark-gluon interaction. Certainly, if one strives for a detailed understanding of the internal structure 
of mesons it is vital to investigate (rather than to model) the effects of the Yang-Mills sector on experimental 
observables much more explicitly. This is one of the two central aims of the present work. Based on previous 
studies \cite{Alkofer:2000wg,Fischer:2002hn} of the pure Yang-Mills sector, results for the quenched and unquenched 
ghost, gluon and quark propagators from their coupled set of \DS equations (SDEs) have been obtained in 
Ref.~\cite{Fischer:2003rp}. Here we extend this setup to include the corresponding Bethe-Salpeter equations and 
determine light pseudoscalar and vector meson observables. The effective interaction of the quark and gluons is 
hereby explicitly resolved in terms of the dressed gluon propagator and the dressed quark-gluon vertex. 

The second aim of this work is to quantify a class of unquenching effects in meson observables. In general, 
'unquenching' means the inclusion of quark 
loops to the various correlation functions that enter the expressions for the quark self-energy and \BS kernel.  A 
previous study of unquenching focused on the mechanism for meson decay \cite{Watson:2004jq} where a single quark 
loop could be incorporated into the truncation scheme, retaining the crucial property of obeying the axialvector 
Ward-Takahashi identity.  However, there exists another viable improvement to the basic \BS truncation scheme and 
this involves the unquenching of the input Yang-Mills interaction form with an infinite series of increasing numbers 
of quark loops. This is the second main issue of the present work.

The paper is organised as follows: In section \ref{sec:SDE} we detail our treatment of the ghost, gluon and quark 
propagators. We investigate unquenching effects due to quark loops in these quantities in two different setups.
On the one hand we explicitly solve the coupled set of (unquenched) \DS equations for the ghost, gluon and quark 
propagators employing ans\"atze for the various vertex functions. On the other hand we investigate the same effects 
employing a model for the quenched Yang-Mills sector of QCD, which we effectively unquench by adding a quark-loop
to the model gluon propagator. The results in both setups are used as input for our calculation of pseudoscalar
and vector meson properties, detailed in section \ref{sec:BSE}. Numerical results and conclusions are reported 
in sections \ref{sec:Results} and \ref{sec:Conclusions}. Various technical aspects are discussed in the appendices.
Throughout the various stages of our calculation we will compare our results from the SDE/BSE approach 
with those from recent lattice simulations. Finally, it is worth noting that whenever we refer to unquenching 
in the present work we only refer to unquenching effects due to the presence of quark loops in the gluon-SDE. 
Certainly, as already mentioned above, unquenching the full theory involves other important effects like meson 
decays which are beyond the scope of this study.


\section{Unquenched propagators}\label{sec:SDE}

To start with, we describe in some detail the two calculational schemes that we use to determine the effects of quark loops 
on the propagators of QCD. 
The first scheme employs the full set of coupled unquenched SDEs for the ghost, gluon and quark
propagators together with appropriate ans\"atze for the ghost-gluon, the three-gluon and the quark-gluon vertex. 
We use a slightly modified truncation compared to the
one used in Ref.~\cite{Fischer:2003rp}. The second scheme employs a model for the quark-gluon-interaction, which has been
shown in Ref.~\cite{Alkofer:2002bp} to be well suited for phenomenological investigations. In both schemes we take
unquenching effects in the gluon polarisation explicitly into account.

\subsection{Propagators from the quark, gluon and ghost \DS equations}\label{subsec:SDE_full}

 For the purpose of our investigation we work in Landau gauge QCD, with the gauge fixing performed by
 the familiar Faddeev-Popov method. The resulting determinant is conveniently represented by a pair of
 Grassmann-fields, the Faddeev-Popov ghosts. The two-point functions of Landau gauge QCD, the ghost, gluon and
 quark propagators can be generically written as\footnote{We work in Euclidean space with scalar product 
 $a\s b=\de_{\mu\nu}a_{\mu}b_{\nu}$.  The Dirac matrices are Hermitian and obey 
 $\{\ga_{\mu},\ga_{\nu}\}=2\de_{\mu\nu}$.}
 \begin{eqnarray}
  D_G^{ab}(p) &=& - \frac{G(p^2)}{p^2}\delta^{ab} \,,
  \label{ghost_prop}\\
  D_{\mu \nu}^{ab}(p) &=& \left(\delta_{\mu \nu} - \frac{p_\mu
      p_\nu}{p^2} \right) \frac{Z(p^2)}{p^2}\delta^{ab} \, ,
  \label{gluon_prop} \\
  S_i(p) &=& \frac{1}{-i \pslash \, A_i(p^2) + B_i(p^2)}
  =  \frac{Z_f^i(p^2)}{-i \pslash + M_i(p^2)}\, .
  \label{quark_prop}
 \end{eqnarray}
 Here $G(p^2)$ denotes the ghost dressing function, $Z(p^2)$ the gluon dressing function, and $A_i(p^2)$ and $B_i(p^2)$
 are the vector and scalar quark self energy of the quark flavors $i \in \{u,d,s\}$. In a somewhat different 
 notation, $M_i(p^2) = B_i(p^2)/A_i(p^2)$ is the 
 quark mass function, whereas $Z_f^i(p^2) = 1/A_i(p^2)$ is called the quark wave function. The corresponding bare 
 tree level propagators,
 $D_G^0(p)$, $D^0_{\mu \nu}(p)$ and $S_i^0(p)$ are given by $Z \equiv G \equiv Z_f \equiv 1$ and $M_i(p^2) = m_i$, where the
 $m_i$ are the respective renormalised quark masses in the Lagrangian of QCD. The color factors of objects in the adjoint 
 representation are given explicitly, whereas the color factors of objects in the fundamental representation are treated 
 implicitly.
 
\begin{figure}[t]
\vspace{0.5cm}
\centerline{
\epsfig{file=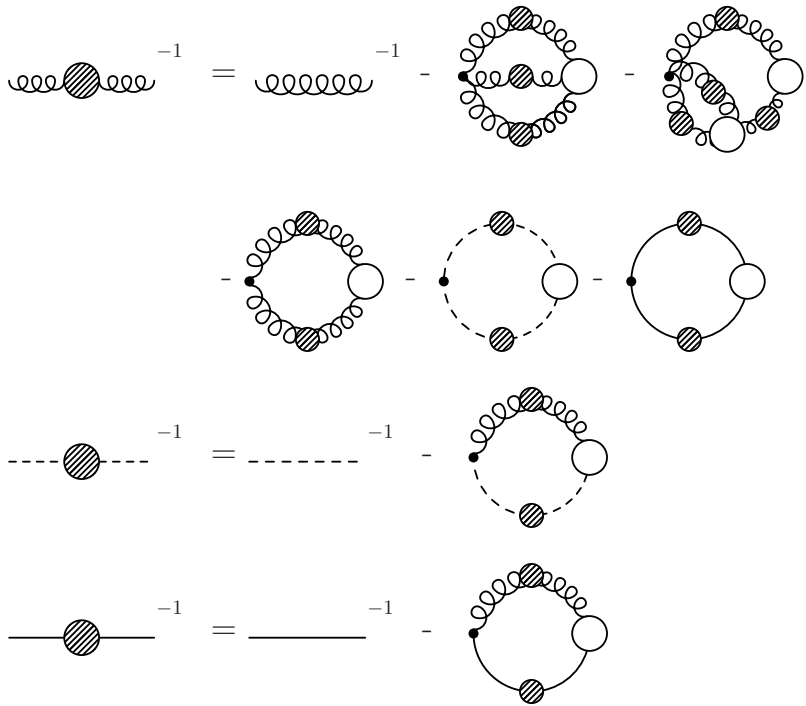,width=0.6\linewidth}
}
\caption{\label{SDEs} A diagrammatical representation of the coupled system of ghost, gluon and quark
\DS equations.  Filled blobs denote dressed propagators and empty circles denote dressed proper vertex functions.
}
\end{figure}

 The propagators (\ref{ghost_prop}), (\ref{gluon_prop}) and (\ref{quark_prop}) satisfy the coupled set of renormalised 
 \DS-equations,
\bea
{[D^{af}(p)]^{-1}_{\mu \nu}}
&=& Z_3 [D^{0,af}(p)]^{-1}_{\mu \nu}  
- \tilde{Z_1}  \int \dk{k} \:\widetilde{\Gamma}^{0,abc}_\mu (p,k) \: 
D_G^{bd}(p-k) \:\widetilde{\Gamma}_\nu^{def}(k,p)D_G^{ce}(k)
\nonumber\\ 
&& - Z_1 \,\frac{1}{2}\,\int \dk{k} 
\:\Gamma^{0,abc}_{\mu \rho \sigma} (p,k) \: 
D_{\rho \rho^\prime}^{bd}(p-k) \:\Gamma_{\rho^\prime \nu \sigma^\prime}^{def}(k,p)
D_{\sigma \sigma^\prime}^{ce}(k) 
\nonumber\\
&& - Z_{1F} \,\frac{1}{2} \, \sum_{i=u,d,s} \int \dk{k} \:\: 
 \mbox{Tr}\left\{ \Gamma_\mu^{0,a} \,S_i(k)\, \Gamma_\nu^a(k,p) \,S_i(p-k)
 \right\} \ + \ \left(\dots\right), 
\label{SDE-gluon} \\
{[D^{af}_G (p)]^{-1}}
&=& \tilde{Z_3} [D_G^{0,ab}(p)]^{-1}  
- \tilde{Z_1} \, \int \dk{k} \:\widetilde{\Gamma}^{0,abc}_\mu (p,k) \: 
D_{\mu \nu}^{bd}(p-k) \:\widetilde{\Gamma}_\nu^{def}(k,p)D_G^{ce}(k) \;, \hspace*{1.5cm} 
\label{SDE-ghost} \\
S_i^{-1}(p) &=& Z_2 \, [S_i^0(p)]^{-1} -  Z_{1F}\,  \int \dk{k} \,
\Gamma_{\mu}^{0,a}(p,k)\, S_i(k) \,\Gamma_\nu^b(k,p) \,D_{\mu \nu}^{ab}(p-k) \,,
\label{SDE-quark}
\eea
which are given diagrammatically in Fig.~\ref{SDEs}.
Here $\dk{k}$ is an abbreviation of $d^4k/(2\pi)^4$. The sum in the quark-loop runs over the light quark flavors 
and the trace is over color and Dirac indices.  The ghost-gluon vertex is denoted by $\widetilde{\Gamma}_\nu^{abc}(k,p)$,
the three-gluon vertex is given by $\Gamma_{\rho \nu \sigma}^{abc}(k,p)$ and ${\Gamma}_\nu^{a}(k,p)$ is the
quark-gluon vertex. The corresponding bare quantities have an additional superscript $0$.
The ellipsis in the gluon-SDE represents two-loop diagrams containing the four-gluon-vertex.
The gluon tadpole diagram vanishes in the process of renormalisation and is therefore omitted from the start.
The SDEs (\ref{SDE-gluon}), (\ref{SDE-ghost}) and (\ref{SDE-quark}) are renormalised and therefore finite
and independent of the ultraviolet regulator $\Lambda$. The definitions of the renormalisation constants 
$Z_i$ are given in appendix \ref{app:A} together with
the corresponding relations between unrenormalised and renormalised dressing functions.

The quenched system of ghost- and gluon-SDEs, Eqs.~(\ref{SDE-ghost}) and (\ref{SDE-gluon}) without the quark loop, 
and the quark-SDE (\ref{SDE-quark}) have been investigated separately in a series of publications 
\cite{vonSmekal:1997is,Atkinson:1997tu,Alkofer:2000wg,Watson:2001yv,Zwanziger:2001kw,Lerche:2002ep,
Fischer:2002hn,Zwanziger:2003cf,Alkofer:2004it,Maris:1997tm,Maris:1997hd,Bloch:2002eq,Bhagwat:2003vw,
Maris:2003vk}. 
Results for the unquenched system, including the back-coupling of the quarks onto the Yang-Mills sector,
have been reported in \cite{Fischer:2003rp}.
In the following we shortly summarise some results, which are important for the present work.

The infrared behaviour of the ghost and gluon propagators can be determined self-consistently from the ghost-SDE
and the ghost-loop in the gluon-SDE alone. This so called ghost-dominance of Yang-Mills theory in the infrared
has been conjectured in Ref.~\cite{vonSmekal:1997is}, lead to the formulation of an infrared effective theory 
based solely on the Faddeev-Popov determinant \cite{Zwanziger:2003cf} and has recently been shown to hold in a 
self-consistent scenario of solutions for general n-point functions \cite{Alkofer:2004it}. In Landau gauge, such
an analysis is based on the factorisation property of the ghost-gluon vertex, which entails that in the presence of
one external scale this vertex cannot develop any nontrivial dressing ({\it i.e.} singularities or zeros) in 
the infrared \cite{Watson:2001yv,Lerche:2002ep,Taylor:ff}. Based on this observation one finds simple power laws,
\begin{equation}
  Z(p^2) \sim (p^2)^{2\kappa}, \hspace*{1cm}  G(p^2) \sim (p^2)^{-\kappa},
  \label{g-power}
\end{equation}
for the gluon and ghost dressing function with exponents related to each other. The relations (\ref{g-power}) 
can be determined from the ghost-SDE alone and are independent of any approximation of the ghost-gluon vertex,
which is in agreement with the factorisation property. The exponent $\kappa$ is an irrational number 
and depends only slightly on the dressing of the ghost-gluon vertex \cite{Lerche:2002ep}. 
With a bare vertex one obtains $\kappa = (93 - \sqrt{1201})/98 \approx 0.595$ \cite{Lerche:2002ep,Zwanziger:2001kw}. 
This result has been confirmed independently in studies of the exact renormalisation group equation \cite{Pawlowski:2003hq}.
Lattice simulations indicate that $\kappa \approx 0.5$ 
\cite{Leinweber:1998im,Bowman:2004jm,Gattnar:2004bf,Oliveira:2004gy,Sternbeck:2005tk}, see however \cite{Fischer:2005ui} for a 
discussion on systematic effects on $\kappa$ due to a compact manifold. 

Probably the most interesting consequence of the power laws (\ref{g-power}) is the corresponding fixed point
of the running coupling $\alpha(p^2)$ in the infrared. A nonperturbative definition of this coupling can be derived
from the ghost-gluon-vertex \cite{vonSmekal:1997is},
\be
\alpha(p^2) = \frac{g^2}{4\pi} G^2(p^2) Z(p^2). \label{coupling}
\ee
No vertex function appears in this definition; a fact that can 
be traced back to the ultraviolet finiteness of the ghost-gluon vertex in Landau gauge. Note that the right hand 
side is a renormalisation group invariant, i.e. $\alpha(p^2)$ does not depend on the renormalisation point.
From the power laws (\ref{g-power}) one finds that the coupling has a fixed point in the infrared, given by \cite{Lerche:2002ep}
\be
\alpha(0) = \frac{4 \pi}{6N_c}
\frac{\Gamma(3-2\kappa)\Gamma(3+\kappa)\Gamma(1+\kappa)}{\Gamma^2(2-\kappa)
\Gamma(2\kappa)} \approx 2.972 \\
\ee
for the gauge group SU(3) and a bare ghost-gluon vertex. Interestingly enough, both, the values of the exponent 
$\kappa$ and the fixed point $\alpha(0)$ can be shown to be independent of the gauge parameter in a class of 
gauges that interpolate between Landau and Coulomb gauge \cite{Fischer:2005qe}. 

Numerical solutions for the ghost and gluon propagator for general momenta have been obtained in refs. 
\cite{Fischer:2002hn,Fischer:2003rp}. In the infrared these solutions agree with the infrared asymptotic behaviour 
of Eq.(\ref{g-power}), whereas in the ultraviolet they reproduce corresponding results from resummed perturbation 
theory. The employed approximation scheme for the ghost-gluon-vertex, the three-gluon-vertex and the four-gluon-vertex
is detailed in \cite{Fischer:2002hn}. For the convenience of the reader we summarise the basic assumptions of this 
truncation in appendix \ref{app:B}. Here we discuss our truncation of the quark-gluon-vertex in more detail, as
this vertex provides the crucial link between the Yang-Mills SDEs, the quark-SDE and the \BS equation describing
mesons as bound states of quarks and antiquarks.

The structure of the quark-gluon-vertex has been the focus of some interest in the last years. It is well known, 
that the leading $\gamma_\mu$-part of the vertex provides sufficient structure to the quark-gluon interaction 
to describe a range of properties of the pseudoscalar and vector mesons well (see \cite{Maris:2003vk} and 
references therein). On the other hand it is also well known by now, that further tensor structure in the vertex,
in particular a scalar part, is mandatory to describe other meson channels. Such corrections to the vertex
have been investigated on a semiperturbative level in several works \cite{Bender:1996bb,Bender:2002as,
Bhagwat:2004hn,Watson:2004kd,Bhagwat:2004kj,Fischer:2004ym} and comparisons with first lattice results 
\cite{Skullerud:2002ge,Skullerud:2003qu} are encouraging (although the error bars from such simulations are 
still quite large). Unfortunately, the numerical complexity of the present investigation is such that 
we are not able to include such corrections. Instead, we will adopt the strategy of Ref.~\cite{Fischer:2003rp}
and employ an ansatz for the nonperurbative dressing of the $\gamma_\mu$-part of the vertex such that
important constraints like multiplicative renormalisability and a correct perturbative limit of the vertex
are satisfied. 

With $k$ denoting the gluon momentum and $p,q$ the two respective quark momenta, two possible ans\"atze for the 
vector part of the quark-gluon-vertex are
\bea
\Gamma_\nu^a(q,p) &=& igT^a \, \gamma_\nu \, \frac{G^2(k^2)}{Z_{1F}} \, A(k^2), \nonumber\\
\Gamma_\nu^a(q,p) &=& igT^a \, \gamma_\nu \, \frac{G(p^2) G(q^2)}{Z_{1F}}\, \frac{A(p^2)+A(q^2)}{2}. \label{quark-gluon-vertex}
\eea
The presence of the vector quark self energy $A(k^2)$ in these expressions is enforced by the Slavnov-Taylor 
identity for the quark-gluon vertex (given in appendix \ref{app:A}), which also accounts for one of the two 
ghost factors. The second ghost factor leads to a perturbative behaviour of the vertex, which produces the correct
perturbative anomalous dimensions of the ghost, gluon and quark propagators in the ultraviolet momentum regime,
provided the momentum arguments of the dressing functions are chosen in an appropriate way. As discussed in
detail in Ref.~\cite{Fischer:2003rp}, the first version of arguments, is necessary in the quark-SDE, whereas
the second version has to be chosen in the quark loop of the gluon-SDE to reproduce the respective correct perturbative
limits of these equations. Via the renormalisation properties of
the contributing dressing functions and the vertex renormalisation constant $Z_{1F}$ this ansatz has the correct 
dependence on the renormalisation scale $\mu^2$ in agreement with the renormalisation group. Furthermore note
that both versions of the vertex have the correct properties under charge conjugation. The price to be paid with
this construction is its cutoff-dependence via $Z_{1F}$. Thus these ans\"atze do not have all properties of well defined
renormalised Green's functions, but have to be regarded as effective constructions, which only make sense in
connection with the quark- and gluon-SDE. If inserted into these equations, the ans\"atze lead to properly 
renormalised, cutoff-independent ghost, gluon and quark propagators.

Together with the truncations for the other vertices, specified in appendix \ref{app:B}, we end up with the following
set of coupled, nonlinear integral equations for the ghost, gluon and quark propagators
\begin{eqnarray} 
\frac{1}{G(p^2)} &=& \tilde{Z}_3 - g^2N_c \int \dk{k}
\frac{K(p^2,k^2,q^2)}{p^2k^2}
G(k^2) Z(q^2) \; , \label{ghost-SDE} \\ 
\frac{1}{Z(p^2)} &=& {Z}_3 + g^2\frac{N_c}{3} 
\int  \dk{k} \frac{M(p^2,k^2,q^2)}{p^2k^2} G(k^2) G(q^2) + 
 g^2 \frac{N_c}{3} \int \dk{k} 
\frac{Q(p^2,k^2,q^2)}{p^2k^2} \frac{G(k^2)^{(-2-6\delta)}}{Z(k^2)^{3\delta}}
\frac{G(q^2)^{(-2-6\delta)}}{Z(q^2)^{3\delta}}     \nonumber\\
&& - g^2 \,  Z_2 \, \sum_{i=u,d,s} \int \dk{k}
\frac{G(k)}{k^2 + M_i^2(k^2)} \frac{G(q)}{q^2 + M_i^2(q^2)} \frac{A_i(k^2)+A_i(q^2)}{2A_i(k^2)A_i(q^2)} W(p^2,k^2,q^2) \,, 
\label{gluon-SDE}\\
B_i(p^2) &=& Z_2 \: m_i + Z_2 \, \frac{4}{3}\, \int \dk{k} \frac{4 \pi \alpha(q^2) A_{u}(q^2)}{q^2} \, 
\frac{3 B_i(k^2)}{k^2 A_i^2(k^2)+B_i^2(k^2)}  \,, \label{B-SDE}\\
A_i(p^2) &=& Z_2 + Z_2 \, \frac{4}{3}\, \int \dk{k} \frac{4 \pi \alpha(q^2) A_{u}(q^2)}{q^2}
\frac{A_i(k^2)}{k^2 A_i^2(k^2)+B_i^2(k^2)} 
\left(-\frac{q^2}{p^2} + \frac{p^2+k^2}{2p^2} + \frac{(p^2-k^2)^2}{2 \: q^2p^2} \right)  \,, \label{A-SDE}
\end{eqnarray} 
where $\delta = -9N_c/(44N_c-8N_f)$ is the one-loop anomalous dimension of the ghost propagator and $i \in \{u,d,s\}$.
Note that with our choice (\ref{quark-gluon-vertex}) of the quark-gluon-vertex it is the running coupling $\alpha(q^2)$ 
(cf. Eq.(\ref{coupling})) multiplied with the quark self energy $A_u(q^2)$ that provides the interaction in the quark-SDEs.
In order to allow the construction of a Bethe-Salpeter kernel in accordance with the axialvector Ward-Takahashi identity 
the interaction in the quark-SDEs has to be flavor independent and is therefore always chosen to contain the $A$-function 
of the lightest (up-)quark.   
The kernels ordered with respect to powers of $q^2=(p-k)^2$ have the form:
\begin{eqnarray}
K(p^2,k^2,q^2) &=& \frac{1}{q^4}\left(-\frac{(p^2-k^2)^2}{4}\right) + 
\frac{1}{q^2}\left(\frac{p^2+k^2}{2}\right)-\frac{1}{4} \,,\nonumber\\
M(p^2,k^2,q^2) &=& \frac{1}{q^2} \left( -\frac{1}{4}p^2 + 
\frac{k^2}{2} - \frac{1}{4}\frac{k^4}{p^2}\right)
+\frac{1}{2} + \frac{1}{2}\frac{k^2}{p^2} - \frac{1}{4}\frac{q^2}{p^2} \,, \nonumber\\
Q(p^2,k^2,q^2) &=& \frac{1}{q^4} 
\left( \frac{1}{8}\frac{p^6}{k^2} + p^4 -\frac{9}{4}p^2k^2 + 
k^4
+\frac{1}{8}\frac{k^6}{p^2} \right)\nonumber\\
&& +\frac{1}{q^2} \left( \frac{p^4}{k^2} - 4p^2-
4k^2+\frac{k^4}{p^2}\right)\nonumber\\
&& - \left( \frac{9}{4}\frac{p^2}{k^2}+4+
\frac{9}{4}\frac{k^2}{p^2} \right) 
+ q^2\left(\frac{1}{p^2}+\frac{1}{k^2}\right) + q^4\frac{1}{8p^2k^2} \nonumber\\
W(p^2,k^2,q^2) &=&\frac{q^4}{3 p^4} + q^2\left(\frac{1}{3p^2} - \frac{2 k^2}{3p^4}\right)
              -\frac{2}{3}+\frac{k^2}{3p^2}  + \frac{k^4}{3q^4} \,.
\label{new_kernels}
\end{eqnarray}
The momentum subtracted versions of Eqs.~(\ref{ghost-SDE})-(\ref{A-SDE}) are employed in our numerical investigations. 
(An outline of the subtraction procedure is given in section \ref{sec:Renorm}.)

\subsection{The phenomenological model}\label{subsec:SDE_model}

As a simpler (at least in some respects) alternative to the above described setup, we employ a model
for the effective interaction between quarks and gluons, which has been developed in \cite{Alkofer:2002bp}.
This effective interaction $\Delta_{\mu \nu}^{ab}$
can be related to the combination of dressed gluon propagator and dressed $\gamma_\mu$-part of
the quark-gluon vertex in the following way:
\be
\Gamma_{\mu}^{0,a}(p,k) \,D_{\mu \nu}^{ab}(p-k) \, \Gamma_\nu^b(k,p) 
\hspace*{5mm} \longrightarrow \hspace*{5mm} \imath T^a\ga_{\mu}\,g^2\Delta_{\mu \nu}^{ab}(p-k)\,\imath T^b\ga_{\nu}
\ee
where both vertices are replaced by their tree-level counterparts ($\G_{\mu}^{0,a}=\imath gT^a\ga_{\mu}$).
The central premise of phenomenologically constructing the effective interaction is that of integrated 
infrared strength -- the interaction, when input into the quark self-energy integral, must be capable of generating 
sufficient amounts of dynamical chiral symmetry breaking [D$\chi$SB] in order to generate the experimentally observed
hadronic mass spectra.  One such construction is the quenched, Landau gauge form
\be
g^2\Delta_{\mu\nu}^{0,ab}(p)=\de^{ab}\,t_{\mu\nu}(p)\,4\pi^2\,d\,\frac{p^2}{\w^2}\,
                      \exp{\left(-\frac{p^2}{\w^2}\right)}\,,
\label{eq:gauss}
\ee
where $t_{\mu\nu}(p)$ is the transverse momentum projector.  The parameters $d$ and 
$\w$ are fitted by comparing the resultant pseudoscalar meson mass and leptonic decay constant to experiment.  
The parameter $\w$, which sets the position of the maximum of the interaction, can be regarded as a length scale associated
with the interaction, and $d$ relates to the overall magnitude.  It turns 
out that the meson masses (and the pseudoscalar and vector leptonic decay constants) are largely independent of $\w$ in 
the range $\w\sim0.2-0.7\mathrm{GeV}$.  As an example $\w=0.5\mathrm{GeV}$, $d=16\mathrm{GeV}^{-2}$ gives realistic 
results for the pseudoscalar and vector channels \cite{Alkofer:2002bp}.
\begin{figure}[t]
\mbox{\epsfig{figure=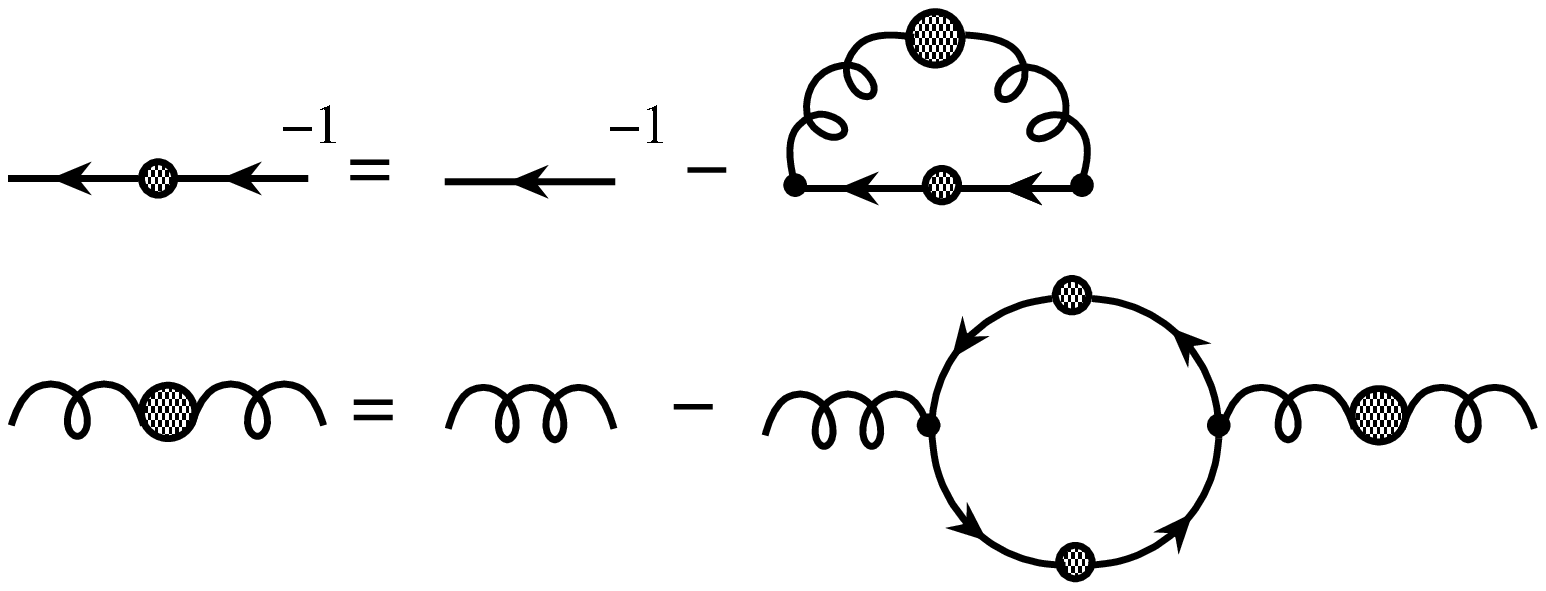,width=13cm}}
\caption{\label{fig:SDE}\DS equations for the quark and gluon propagator under the truncation scheme of the 
phenomenological model.  All vertices are tree-level.  Blobs represent fully dressed propagators.  The seed 
term of the gluon equation is dressed with the Gaussian form Eq.~\ref{eq:gauss} (see text).}
\end{figure}
Clearly, by fitting to experiment, there is implicit information contained within such a phenomenological interaction 
about quark loops contained within the choice of parameters.  However, the quark loops are at least from the viewpoint 
of truncated equations an additional degree of freedom.  In the context of a full field theory (QCD here) such inclusions 
are self-consistently present as described in the previous subsection. In order to make such contributions also apparent
in this phenomenological model we unquench the effective interaction by adding to the phenomenological Gaussian form 
a quark loop term which generates an infinite series of increasing numbers of quark loops. The quark-gluon vertex itself 
remains undressed.  The truncated set of SDEs for this phenomenological model is shown in Fig.~\ref{fig:SDE}.

The equation for the unquenched effective interaction is
\be
\De_{\mu\nu}^{ab}(p)=\De_{\mu\nu}^{0,ab}(p)-\,\De_{\mu\la}^{0,ac}(p)\, g^2\de^{cd}\, t_{\la\ro}(p)\, p^2\,\Si(p^2)\,
\De_{\ro\nu}^{db}(p)
\label{eq:eff}
\ee
with
\be
g^2\,\de^{cd}\,t_{\la\ro}(p)\,p^2\,\Si(p^2)=g^2\,\de^{cd}\,t_{\la\ro}(p)\,p^2\,Z_{\De}+\int \dk{k} \sum_{i=u,d,s} \, 
Tr\left[\imath gT^c\ga_{\la}\,S^i(k)\,\imath gT^d\ga_{\ro}\,S^i(k-p)\right]
\label{eq:pol}
\ee
where the sum runs over the light quark flavors and the trace is over color and Dirac indices.  
The prefactors in the definition of $\Si$ are such that $\Si$ is a dimensionless scalar function.  
The renormalisation constant $Z_{\De}$ is independent of the external momentum scale and has been introduced to cancel 
the UV divergence of the quark loop.
One can see that the integral of the right-hand side in \eq{eq:pol} is transverse with respect to momentum $p$ in exactly 
the same way as for the one-loop perturbative quark loop contribution to the gluon polarisation due to the presence of 
the tree-level vertices. Making the color and tensor structure of the effective interaction explicit with
\be
\De_{\mu\nu}^{ab}(k)=\de^{ab} t_{\mu\nu}(k) \De(k^2)
\ee
and projecting out the scalar part $\De(k^2)$ from Eqs.~(\ref{eq:eff}) and (\ref{eq:pol}) (see section \ref{sec:Renorm} for 
more discussion on the projection) we arrive at
\be
\De(p^2)=\frac{\De^0(p^2)}{1+g^2\De^0(p^2) p^2 \Si(p^2)}
\label{eq:int1}
\ee
with
\be
\Si(p^2) =Z_{\De}+ \frac{4}{3} \, \int\dk{k} \, \sum_{i=u,d,s} \, \frac{1}{k^2 + M^2(k^2)} 
\frac{1}{q^2 + M^2(q^2)} \frac{1}{A(k^2)A(q^2)} 
\left[\frac{k^2}{p^2}-4\frac{p\s k^2}{p^4}+3\frac{p\s k}{p^2}\right].
\label{eq:s1}
\ee
where the momentum $q=k-p$.

For the quark dressing functions $A(p^2)$ and $B(p^2)$ we obtain
\bea
B(p^2) &=& m + \frac{4}{3}\,\int \dk{k} g^2\De(q^2) \, \frac{3 B(k^2)}{k^2 A^2(k^2)+B^2(k^2)}  \,, \label{eq:b1}\\
A(p^2) &=& 1 + \frac{4}{3}\,\int \dk{k} g^2\De(q^2) \, \frac{A(k^2)}{k^2 A^2(k^2)+B^2(k^2)} 
\left(-\frac{q^2}{p^2} + \frac{p^2+k^2}{2p^2} + \frac{(p^2-k^2)^2}{2 \: q^2p^2} \right)  \,, \label{eq:a1}
\eea
Note that the renormalisation constants of the quark equations above have been set to unity since the effective interaction is 
exponentially damped in the UV and the subsequent integrals are convergent.
These last four equations are used for our numerical investigations of unquenching effects in the phenomenological model. 
Further technical details concerning our method to solve these equations in the complex momentum plane (necessary to solve 
the \BS equation in Euclidean space) are given in appendix \ref{app:C}. 

\section{The Bethe-Salpeter Equation}\label{sec:BSE}

The homogeneous Bethe-Salpeter equation [BSE] for flavor non-singlet quark-antiquark mesons can be written
\be
\G_{\al\ba}^{ij}(p;P)=\int\dk{k}K_{\al\ba;\de\ga}(p,k;P)\left[S^i(k_+)\G^{ij}(k;P)S^j(k_-)\right]_{\ga\de}
\label{eq:bse}
\ee
where $K$ is the Bethe-Salpeter kernel, momenta $k_+=k+\xi P$ and $k_-=k+(\xi-1)P$ are such that the total momentum 
$P=k_+-k_-$ and $\xi=[0,1]$ is the momentum partitioning parameter reflecting the arbitrariness in the relative momenta 
of the quark-antiquark pair.  The flavor content is expressed through the indices $i,j$ and for flavor non-singlet mesons 
the kernel can be written independently of flavor content.  Greek indices ($\al\ldots$) 
refer to color and Dirac structure.  The BSE is a parametric eigenvalue equation with discrete solutions 
$P^2=-M_n^2$ where $M_n$ is the mass of the resonance.  The lowest mass solution corresponds to the physical ground state.  
Since $P^2$ is negative, the momenta $k_{\pm}$ are necessarily complex in Euclidean space and so the quark propagator functions 
 must be evaluated with complex argument.  This leads to technical issues that are dealt with in appendix \ref{app:C}. 

One of the defining characteristics of the light pseudoscalar meson sector is the pattern of chiral symmetry breaking 
and the identification of pions as the Goldstone bosons of the symmetry breaking.  In the Bethe-Salpeter formalism this 
is associated with observance of the axialvector Ward-Takahashi identity [AXWTI] \cite{Maris:1997hd}.  The AXWTI serves 
to relate the quark self-energy and the Bethe-Salpeter kernel such that the pattern of chiral symmetry breaking in the 
pseudoscalar meson sector is guaranteed, independently of truncation or model details.  In the flavor non-singlet channel 
and (for illustrative purposes) with equal mass quarks the AXWTI is
\be
-\imath P_{\mu}\G_{\mu}^5(p;P)=S^{-1}(p_+)\ga_5+\ga_5S^{-1}(p_-)-2m_r\G^5(p;P)
\ee
where we have introduced the (color-singlet) axialvector and pseudoscalar vertices ($\G_{\mu}^5$ and $\G^5$, respectively) 
and $m_r$ is the renormalised quark mass.  Defining
\be
\La(p;P)=-\imath P_{\mu}\G_{\mu}^5(p;P)+2m_r\G^5(p;P)
\ee
then the inhomogeneous \BS integral equations for $\G_{\mu}^5$ and $\G^5$ can be combined to give (replacing the color 
and Dirac indices)
\be
\La_{\al\ba}(p;P)=\left[\imath Z_2\ga_5\slr{P}+Z_42m_r\ga_5\right]_{\al\ba}+\int\dk{k}K_{\al\ba;\de\ga}(p,k;P)
\left[S(k_+)\La(k;P)S(k_-)\right]_{\ga\de}.
\ee
Using the AXWTI to rewrite $\La$ in terms of the quark propagator gives
\be
\left[S^{-1}(p_+)\ga_5+\ga_5S^{-1}(p_-)\right]_{\al\ba}=\left[\imath Z_2\ga_5\slr{P}+Z_42m_r\ga_5\right]_{\al\ba}
+\int\dk{k}K_{\al\ba;\de\ga}(p,k;P)\left[\ga_5S(k_-)+S(k_+)\ga_5\right]_{\ga\de}
\ee
and by writing the \DS equation for the quark as
\be
S^{-1}(p)=-Z_2\imath\slr{p}+Z_4m_r-\Si_f(p)
\ee
one can eliminate the inhomogeneous terms to get
\be
\left[-\Si_f(p_+)\ga_5-\ga_5\Si_f(p_-)\right]_{\al\ba}=\int\dk{k}K_{\al\ba;\de\ga}(p,k;P)
\left[\ga_5S(k_-)+S(k_+)\ga_5\right]_{\ga\de},
\label{eq:axwti}
\ee
thereby making the relationship between the quark self-energy $\Si_f$ and the \BS kernel $K$ manifest.

For the truncation scheme given by the full-SDE set, in order to satisfy \eq{eq:axwti} we specify the truncated \BS kernel to be:
\be
K_{\al\ba;\de\ga}(p,k;P)\rightarrow \left[\Gamma_{\mu}^{0,a}(p_+,k_+)\right]_{\al\ga}
\left[\Gamma_\nu^{0,b}(k_-,p_-)\right]_{\de\ba} \de^{ab} t_{\mu\nu}(p-k) Z_2 \frac{4\pi}{g^2}\frac{\alpha(p-k) A_u(p-k)}{(p-k)^2}\,,
\ee
where $A_u$ is the vector self energy of the up-quark regardless of the remaining flavor structure, which is left implicit.
In the case of the phenomenological model, the kernel is nothing more than the standard dressed ladder truncation:
\be
K_{\al\ba;\de\ga}(p,k;P)\rightarrow \left[\Gamma_{\mu}^{0,a}(p_+,k_+)\right]_{\al\ga}
\left[\Gamma_\nu^{0,b}(k_-,p_-)\right]_{\de\ba}\de^{ab} t_{\mu\nu}(p-k) \De(p-k).
\ee
Notice that the addition of the explicit quark loop in the expression for the interaction does not alter the structure of the 
kernel. Expanding out the various factors (retaining the flavor content), the \BS equation for the two schemes are written as
\be
\G^{ij}(p;P)=-\frac{4}{3} Z_2 \int\dk{k} \frac{4\pi \alpha(p-k) A_u(p-k)}{(p-k)^2} t_{\mu\nu}(p-k)
              \ga_{\mu}S^i(k_+)\G^{ij}(k;P)S^j(k_-)\ga_{\nu}
\ee
for the full SDE scheme and
\be
\G^{ij}(p;P)=-\frac{4}{3}\int\dk{k} g^2 \De(p-k)t_{\mu\nu}(p-k)\ga_{\mu}S^i(k_+)\G^{ij}(k;P)S^j(k_-)\ga_{\nu}
\ee
for the phenomenological model.

Meson channels are specified by the quantum numbers $J^{PC}$ and flavor content.  We assume isospin degeneracy ($m_d=m_u$) 
and consider only flavor non-singlet channels such that the flavor content of the meson can be generated in the BSE by 
supplying the appropriate quark propagators $S^i$ (henceforth we drop the flavor indices).  The $J^{PC}$ quantum 
numbers are input into the BSE by specifying the Dirac and Lorentz decomposition of the bound-state vertices $\G$ 
to have the relevant transformation properties.  We shall be considering pseudoscalar ($0^{-+}$) and vector ($1^{--}$) 
mesons whose decompositions are:
\bea
\G^{PS}(p;P)&=&\ga_{5}\left[\G_0(p;P)-\imath\slr{P}\G_1(p;P)-\imath\slr{p}
\G_2(p;P)-\left[\slr{P},\slr{p}\right]\G_3(p;P)\right],\label{eq:pseu}\\
\G_{\mu}^{V}(p;P) &=& \ga_{\mu}^T \left[
\imath\G_0(p;P)+\slr{P}\G_{1}(p;P)-\slr{p}\G_2(p;P)+\imath\left[\slr{P},
\slr{p}\right]\G_3(p;P)\right]\nonumber\\
&&+ p_\mu^T \left[\G_{2}(p;P)+2\imath
\slr{P}\G_3(p;P) \right]\nonumber\\
&&+ p_\mu^T \left[\G_4(p;P)+
\imath\slr{P}\G_5(p;P)-\imath\slr{p}\G_6(p;P)+\left[\slr{P},\slr{p}\right]
\G_7(p;P)\right],\label{eq:vect}
\eea
where $\ga_{\mu}^T$ and $p_\mu^T$ are transverse to $P_\mu$ (the on-shell vector meson is transverse to its total momentum $P$).  
The $\G_i$ are scalar functions of $p^2$ and $p\s P$ ($P^2$ is on-shell for solutions of the BSE and therefore fixed).  
The charge conjugation properties of the decompositions above are such that when considering 
charge eigenstates, the $\G_i$ are either odd or even under the interchange $p\s P\rightarrow-p\s P$.  By projecting 
the BSE onto the covariant basis it becomes a coupled set of nonlinear integral equations for the $\G_i$.
The technical procedure of solving these equations is outlined in section \ref{sec:solveBSE}. 

The solution of the pseudoscalar BSE can be used to obtain the pion leptonic decay constant, $f_{\pi}$, which will 
prove useful in fitting parameters to observables.  In order to calculate $f_{\pi}$ one must firstly normalise the 
BS vertex function $\G$.  The normalisation condition is derived by demanding that the residue of the pole in the 
four-point quark-antiquark Green's function (from which the BSE is derived) be unity 
\cite{Tandy:1997qf,Nakanishi:1969ph}.  For $\xi=1/2$ it reads
\be
2P_{\mu}=3Tr_d\int\dk{k}\left\{\ov{\G}(k,-P)\frac{\partial S(k+P/2)}{\partial P_{\mu}}
\G(k,P)S(k-P/2)+\ov{\G}(k,-P)S(k+P/2)\G(k,P)\frac{\partial S(k-P/2)}{\partial P_{\mu}}\right\}_{P^2=-M^2}
\ee
where the trace is over Dirac matrices, both propagators describe $u$-quarks and the conjugate vertex function 
$\ov{\G}$ is defined as
\be
\ov{\G}(p,-P)=C\G^T(-p,-P)C^{-1}
\ee
with the charge conjugation matrix $C=-\ga_2\ga_4$.  The leptonic decay constant characterising the pion coupling 
to the point axial field is subsequently given by \cite{Tandy:1997qf}
\be
f_{\pi}=\frac{3}{M^2}Tr_d\int\dk{k}\G(k,-P)S(k+P/2)\ga_5\slr{P}S(k-P/2),
\label{eq:fpi}
\ee
where again the trace is over Dirac matrices.  Eq.~(\ref{eq:fpi}) can also be applied to the kaon by inserting the corresponding 
quark propagators (with the apropriate normalisation).


\section{Numerical Procedure and Results}\label{sec:Results}

\subsection{Technical details}
\subsubsection{Renormalisation and subtraction procedure} \label{sec:Renorm}

To regularise the coupled system of ghost, gluon and quark SDEs we apply a MOM regularisation 
scheme \cite{vonSmekal:1997is,Atkinson:1997tu,Kizilersu:2001pd}. If we write the 
equations (\ref{ghost-SDE}), (\ref{gluon-SDE}), (\ref{B-SDE}) and (\ref{A-SDE}) symbolically as
\bea
\frac{1}{G(p^2)} &=& \tilde{Z}_3 + \Pi_{ghost}(p^2) \,, \\
\frac{1}{Z(p^2)} &=& {Z}_3 + \Pi_{glue}(p^2) \,, \\
A(p^2) &=& Z_2 + Z_2 \, \Pi_A(p^2) \,, \\
B(p^2) &=& Z_4 \, m_R  + Z_2 \, \Pi_B(p^2) \,, 
\eea
where $m_R$ is the renormalised quark mass in the Lagrangian of the theory,
this procedure yields
\bea
\frac{1}{G(p^2)} &=& \frac{1}{G(\mu^2)} + \Pi_{ghost}(p^2) - \Pi_{ghost}(\mu^2) \,, \\
\frac{1}{Z(p^2)} &=& \frac{1}{Z(\mu^2)} + \Pi_{glue}(p^2)  - \Pi_{glue}(\mu^2) \,, \\
\frac{1}{A(p^2)} &=& 1 - \frac{1}{A(p^2)}\Pi_A(p^2) + \Pi_A(\mu^2) \,, \\
B(p^2)           &=& B(\mu^2) + \frac{A(\mu^2)}{1+\Pi_A(\mu^2)} \, \left(\Pi_B(p^2) - \Pi_B(\mu^2) \right) \,,
\eea
where $\mu^2$ is a suitable renormalisation point.
The a priori unknown renormalisation constants $Z_3$, $\tilde{Z}_3$ and $Z_2$ drop out and instead  
one specifies normalisation conditions for $G(\mu^2)$, $Z(\mu^2)$ and $A(\mu^2)$.
For a given renormalised coupling $g$ the Slavnov-Taylor identity
\be
\label{sti}
1=\widetilde{Z}_{1} = Z_g\: \widetilde{Z}_3\: Z_3^{1/2}.
\ee
enforces that one cannot choose the normalisation conditions for $G(\mu^2)$ and $Z(\mu^2)$ separately,
but only for the product $Z(\mu^2)\,G^2(\mu^2)$, see Ref.~\cite{vonSmekal:1997is}. Multiplicative renormalisability 
implies that the dressing functions for different choices of $\mu$ are trivially related to each other
by factors of the renormalisation constants. Note that the coupling $\alpha(p^2)=\alpha(\mu^2) G^2(p^2) Z(p^2)$
and the quark mass function $M(p^2)=B(p^2)/A(p^2)$ are independent of the renormalisation point. 

A further complication is the appearance of spurious quadratic divergencies in the gluon-SDE, which are not eliminated
by the MOM procedure. In principle, quadratic divergencies can be avoided by contracting 
the right hand side of the gluon-SDE with the Brown-Pennington projector 
$R_{\mu\nu}(p)=\de_{\mu\nu}-4\frac{p_{\mu}p_{\nu}}{p^2}$ \cite{Brown:1988bm,Brown:1988bn}. However, using this 
technique one picks up spurious longitudinal contributions from the right hand side of the
gluon-SDE, which affect the value of $\kappa$ in the infrared power laws of the ghost and gluon dressing function, 
Eqs.~(\ref{g-power}) \cite{Lerche:2002ep,Fischer:2002hn}.
We therefore employ an alternative procedure, which is well known in the context of 
exact renormalisation group equations \cite{Pawlowski:2003hq,Ellwanger:1996wy}. After contracting the right hand 
side of the gluon-SDE with the transverse projector, leading to Eq.~(\ref{gluon-SDE}), 
quadratic divergencies show up as terms proportional to $\Lambda^2/p^2$, where $\Lambda$ is the UV-cutoff.
In the infrared, such terms can be identified unambiguously by fitting the right hand side to the form $a/p^2 + b/(p^2)^{2\kappa}$.
The coefficient $a$ then measures all contributions from quadratic divergencies and is subtracted from the right hand side at
each step of the numerical iteration procedure. Together with the MOM-scheme described above this procedure ensures that
all renormalised dressing functions are independent of the regularisation scale\footnote{In 
Ref.~\cite{Fischer:2002hn} quadratic divergencies have been eliminated by a different, formally less rigorous procedure, 
which nevertheless leads to the same results as the present approach within numerical accuracy.}.
 
The situation is somewhat simpler for the phenomenological model, specified in section \ref{subsec:SDE_model}.  
The projection of the quark loop contribution to the interaction Eq.~(\ref{eq:pol}) can be carried out using the Brown-Pennington 
projector to eliminate quadratic divergences without further complication.  
Also, the form of the effective interaction \eq{eq:int1} with the Gaussian term $\De^0(p^2)$ alters the picture.  
$\De^0(p^2)$ (and hence $\De(p^2)$) vanishes exponentially in the UV, leaving the integrals for \eq{eq:a1} and \eq{eq:b1} 
UV convergent.  The remaining integral for $\Si(p^2)$ does 
though require regularisation and renormalisation since the quark functions do reduce to their tree-level values in the UV, 
producing a logarithmic divergence. Also here we use a subtractive scheme and define the renormalised dressing function 
$\ov{\Si}(p^2)$ to be
\be
\ov{\Si}(p^2)=\Si(p^2)-\Si({\mu^2})
\ee
such that $\ov{\Si}(p^2=\mu^2)=0$ for some spacelike renormalisation scale $\mu$.  We then take as our interaction
\be
\De(p^2)=\frac{\De^0(p^2)}{1+g^2\De^0(p^2) \,p^2 \,\ov{\Si}(p^2)}.
\label{eq:int2}
\ee
In what follows, we drop the overbar on $\Si$ and the subtraction is implicitly assumed though not written explicitly for 
notational convenience.  The renormalisation scale $\mu$ is in principle a free parameter of the model.  However, since we are 
studying light mesons with an associated hadronic scale of $\sim1\mathrm{GeV}$ we expect $\mu$ to be in this region.  It turns out 
that there is only a slight dependence of the overall results on $\mu$, reflecting the approximate observance of renormalisation 
scale invariance (c.f. appendix \ref{app:D}). Throughout the main body of the paper we use the single value $\mu=1\mathrm{GeV}$.

There is one major distinction between the full SDE set and the phenomenological model and this concerns the role of the  
quark mass $m_R$.  In the full set of SDEs, the renormalisation of the quark mass and its subsequent running with momentum scales 
allows one to compare the current quark mass with the perturbative quark mass.  In the case of the phenomenological model 
however, the quark mass is nothing more than a parameter of the model and cannot be directly compared to any other quantity.

\subsubsection{Solving the Bethe-Salpeter equation} \label{sec:solveBSE}

Projecting the BSE, Eq.~\ref{eq:bse} onto the covariant basis $\G_i$ 
(for pseudoscalar and vector mesons given via the decompositions Eqs.~\ref{eq:pseu} and \ref{eq:vect} respectively) 
it becomes a coupled set of 
nonlinear integral equations. To solve this set the $\G_i$ are approximated by a Chebyshev expansion in the angular variable 
$z=p\s P/\sqrt{p^2P^2}$,
\be
\G_i(p^2,z;P^2)=\sum_{k=0}^{N_{ch}-1}\imath^kT_k(z)\G_i^k(p^2;P^2).
\ee
Since the $T_k$ form an orthonormal set one can further project the equations onto this basis.  In the case of charge 
eigenstate mesons, one can extract an explicit factor $p\s P$ from the odd $\G_i$ and then use only the even $T_k$.  
The meson mass solution to the system of coupled equations in the $\G_i^k$ is then found in one of two equivalent ways.  
The first is to introduce a fictitious linear eigenvalue $\la(M)$ so that the original BSE reads
\be
\la(M)\G_{\al\ba}(p;P)=\int\dk{k}K_{\al\ba;\de\ga}(p,k;P)\left[S(k_+)\G(k;P)(k_-)\right]_{\ga\de}
\ee
with $P^2=-M^2$ and vary the mass until $\la(M=M_n)=1$ (the physical ground state corresponds to the largest real 
eigenvalue).  The second method is to recognise that after decomposition, the final integral over the radial momentum 
must be discretized giving rise to a matrix equation 
$\G_i^k(p_j^2;P^2)=K_{ii'}^{kk'}(p_j^2,p_{j'}^2;P^2)\G_{i'}^{k'}(p_{j'}^2;P^2)$ whose solution is the highest 
(physical ground state mass) $P^2=-M^2$ for which $\mbox{Det}(1-K)=0$.

In principle, a truncation of the Chebyshev expansion destroys the Poincar\'e covariance of the results, expressed through 
the invariance of the mass solution $M^2$ with the momentum sharing parameter $\xi$.  However, the expansion quickly 
converges such that the numerical results are explicitly stable with varying $\xi$ \cite{Alkofer:2002bp}.  In this work we take 
$\xi=1/2$ and $N_{ch}=4(6)$ for the pseudoscalar (vector) meson channels.

\subsection{Numerical Results for the Propagator Functions}

\begin{figure}[t]
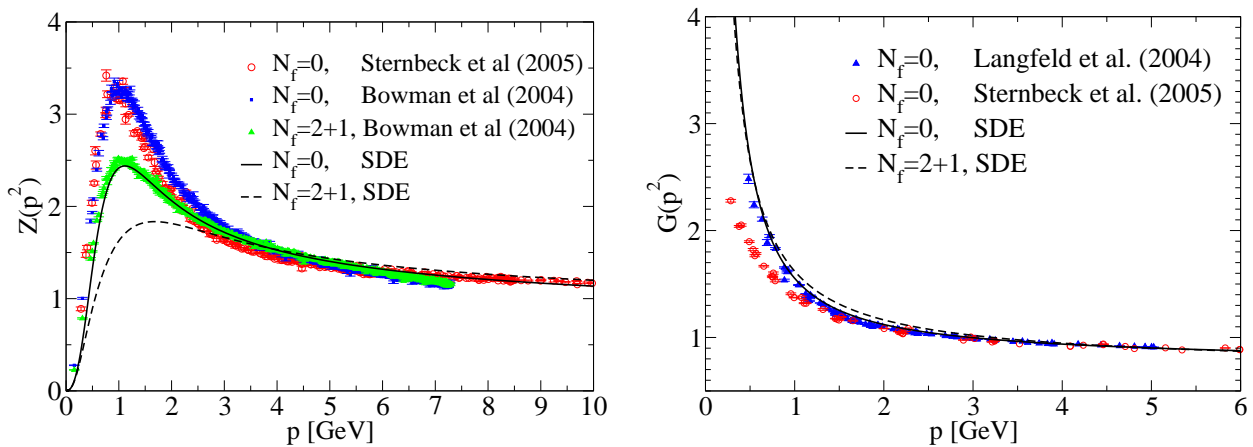

\mbox{\epsfig{figure=GGA.gluelatt.eps,width=8cm}\hspace{0.5cm}\epsfig{figure=GGA.ghostlatt.eps,width=8cm}}\\
\caption{\label{fig:YM-dress}Comparison of the quenched and unquenched ghost and gluon dressing functions
with recent lattice data \cite{Sternbeck:2005tk,Bowman:2004jm,Gattnar:2004bf}. 
The sea-quark masses are $m_{u/d} \simeq 16 \mbox{MeV}, m_s \simeq 79 \mbox{MeV}$ in the lattice
simulations and  $m_{u/d} \simeq 3.9 \mbox{MeV}, m_s \simeq 84 \mbox{MeV}$ in the SDE-approach, c.f. table \ref{tab:fpar}.}
\end{figure}

Our results for the (un)quenched ghost and gluon dressing functions from the full set of SDEs are displayed in 
figure~\ref{fig:YM-dress}. In the infrared, 
the numerical SDE-results reproduce the analytical power laws, Eqs.(\ref{g-power}).  (This can be seen explicitly on a 
log-log-plot, displayed e.g. in Ref.~\cite{Fischer:2002hn}). In the ultraviolet they reproduce the correct one-loop running of 
resummed perturbation theory, with anomalous dimensions $\gamma = \frac{-13 N_c + 4 N_f}{22 N_c - 4 N_f}$ and 
$\delta = \frac{-9 N_c}{44 N_c - 8 N_f}$ for the gluon and ghost dressing functions. Compared to the results of recent lattice 
calculations \cite{Sternbeck:2005tk,Bowman:2004jm,Gattnar:2004bf} we find good agreement for large and small 
momenta\footnote{Deviations for the value of the infrared exponent $\kappa$ may have systematical reasons, as discussed in 
\cite{Fischer:2005ui}. These deviations are not important for the present investigation.}. In the intermediate momentum region one 
clearly sees unquenching effects in the gluon dressing function. In this momentum region the system has enough energy to create 
quark-antiquark pairs from the vacuum. The screening effect from these pairs, first seen in Ref.\cite{Fischer:2003rp},
drives the gluon dressing function closer to its perturbative form, {\it i.e.} the bump in the dressing functions becomes 
smaller. This effect is pronounced in both, the SDE and the lattice results. The overall difference in the size of the 
gluon bump between the two approaches indicates that the (neglected) gluon-two-loop diagrams in the gluon-SDE have an influence
in this momentum region. Moreover, adding additional tensor structure to the quark-gluon vertex changes the magnitude of the
unquenching effects slightly, as can be seen by comparison with Fig.1 in Ref.~\cite{Fischer:2004wf}.
In general, however, we expect that the inclusion of these contributions (in future work) will not change any of the 
conclusions drawn in this paper. 

In comparing the quenched and unquenched interactions of the full SDE set and the phenomenological model, the situation is 
complicated by the presence of non-trivial quark masses.  As pointed out previously, in the phenomenological model the quark 
masses are no more than parameters to be fitted to experimental results for meson observables (see next subsection for details).  
They cannot be compared directly to those of either the full SDE set or experiment.  In addition, for both cases, the back 
reaction of the unquenching term induces changes to the fitted quark masses which further obscures the comparison.  
Therefore, to start the discussion we restrict for now to the chiral limit where a more unambiguous comparison can be made.  
Using the spacelike chiral quark propagator alone there are two quantities that can be compared in the quenched and 
unquenched truncations -- the vacuum chiral quark condensate $<\ov{q}q>^0$ and the quark mass function at zero momentum, 
$M(0)$. The condensate is the order parameter for dynamical chiral symmetry breaking and at a renormalisation 
scale $\mu$ is defined as (see for example Ref.~\cite{Langfeld:2003ye} for a discussion on this topic)
\be
<\ov{q}q>^0(\mu)=-\lim_{\La\rightarrow\infty}Z_4(\mu,\La)N_cTr_d\int\dk{k}S_0(k)
\label{eq:cond}
\ee
where $\La$ is the regularisation scale, $S_0$ is the chiral limit of the quark 
propagator and the trace is over Dirac matrices. In the full set of SDEs we determine the (un)quenched condensate from
the chiral quark propagator renormalised at a large scale $\mu^2$ and subsequently evolve the value down to 
$\mu=2\mathrm{GeV}$ employing one-loop scaling.  
For the phenomenological model, where $Z_4=1$, the condensate depends only indirectly on $\mu$ via the unquenching effects.
In both approaches the unquenching effects on the quark-gluon interaction are calculated with physical
sea quark configurations (see the next subsection for details of the fitting procedure and parameter values). 

Our results for $<\ov{q}q>^0$ and $M(0)$ are presented in table \ref{tab:cond_M}.
In the full SDE-setup with the quark-gluon vertex (\ref{quark-gluon-vertex}) we find a very slight enhancement of the condensate 
once quark-loops are taken into account. An even smaller effect has been found in Ref.~\cite{Fischer:2003rp}, where different 
approximations of the vertex have been investigated. These results together indicate that the condensate is almost constant  
for $N_f \le 3$. Since the condensate is an order parameter for D$\chi$SB it is expected to change rapidly at the 
vicinity of the chiral phase transition. We conclude therefore that the critical number of flavors for the
chiral transition of QCD at zero temperature is much larger than $N_f = 3$. This agrees well with the recently estimated 
value $N_f^c = 10.0\pm 0.4$ in the framework of exact renormalisation group equations \cite{Gies:2005as} and expectations
from perturbation theory.  
For the quark mass at zero momentum we note a reduction of $\sim10 \%$ due to quark-loop effects in both the full SDE and 
the phenomenological frameworks. A similar reduction has been found in Ref.~\cite{Fischer:2003rp} for a range of 
different truncations of the quark-gluon vertex. We therefore believe this effect to be model independent. Recent lattice 
simulations using staggered quarks confirm this result \cite{Bowman:2005vx}. 

\begin{table}[t]
\caption{\label{tab:cond_M} The chiral condensate and the chiral quark mass function at $p^2=0$ in the full-SDE setup and in the 
phenomenological model. The unquenching effects are calculated with physical sea-quarks, see tab.~\ref{tab:fpar} for the
corresponding values of the quark masses and interaction parameters.}
\begin{center}\begin{tabular}{|c|c|c|}\hline
& $-(<\ov{q}q>^0)^{1/3}$ ($\mathrm{MeV}$) & $M_{ch}(p^2=0)$ ($\mathrm{MeV}$)\\\hline\hline
Model:    &   &     \\\hline
$N_f=0$   &244& 366   \\\hline
$N_f=2+1$ &243& 332   \\\hline\hline
full SDE: &   &     \\\hline
$N_f=0$   &266& 416 \\\hline
$N_f=2+1$ &271& 388 \\\hline
\end{tabular}\end{center}\end{table}

\begin{figure}[t]
\epsfig{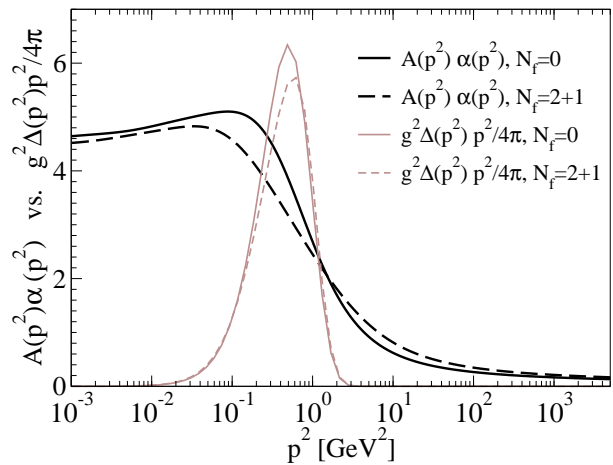}
\caption{\label{fig:alpha-delta}Comparison of the quenched and unquenched interactions used in the two models. 
In the full-SDE framework the interaction is given by the product $A(p^2) \al(p^2)$ whereas for the phenomenological model 
the equivalent function is $g^2\De(p^2)p^2/4\pi$. The unquenching effects are calculated with physical sea-quarks, 
see tab.~\ref{tab:fpar} for the corresponding values of the quark masses and interaction parameters. }
\end{figure}

Next we compare the effective interaction of the phenomenological model, $g^2\De(p^2)\, p^2/4\pi$, to the interaction of the 
full-SDE setup, $\al(p^2)\, A(p^2)$, at finite quark masses. The parameters of the two schemes are fixed to reproduce the physical
pion and kaon observables (see the next subsection for details of the fitting procedure and parameter values).
From the plot in Fig.~\ref{fig:alpha-delta} one notes drastic differences. 
The interaction of the full-SDE setup stretches from the deep infrared to the ultraviolet, where the perturbative one-loop 
running is reproduced. In the infrared the interaction is not vanishing due to the fixed point behaviour of the running 
coupling\footnote{Note that the infrared behaviour of the running 
coupling (defined from the Bjorken sum rule) has been extracted very recently from experimental data at JLAB \cite{Deur:2005cf}
and found to be in excellent agreement with our SDE-result, taken from \cite{Fischer:2002hn}. Since the exact theoretical relation 
between an invariant charge from experimental data and the coupling from the ghost-gluon vertex has not yet been clarified, 
such a comparison has to be treated with caution. Nevertheless, it suggests that the full SDE-interaction reflects essential
properties of the strong interaction at large and small momenta.
}, Eq.~(\ref{coupling}), multiplied with a finite value of $A(0)$. 
The effective interaction of the phenomenological model, however, is localised at 
intermediate momenta. Despite the clear differences, both interactions produce similar qualitative effects in meson 
observables as will be seen in later sections.  This is a validation of the assertion that the simplest $q\bar{q}$ meson 
states are dominated by the pattern of dynamical chiral symmetry breaking which arises from the integrated strength of the 
interaction and are not largely affected by the detailed structure of the interaction itself.

\begin{figure}[t]
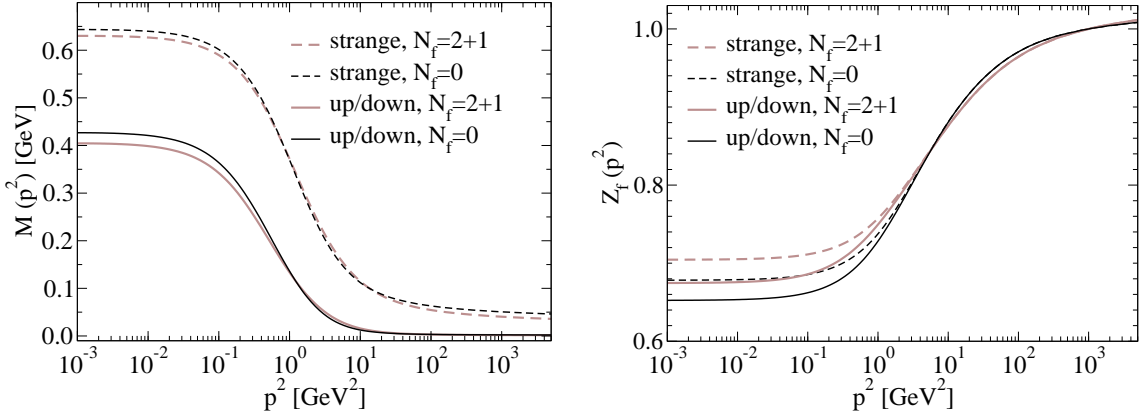

\vspace{1cm}
\mbox{\epsfig{figure=GGA.M.lin.eps,width=7.3cm}\hspace{0.5cm}\epsfig{figure=GGA.A.eps,width=7.3cm}}\\
\caption{\label{fig:qspace_full}Comparison of the quenched and unquenched quark mass and wave-functions (left and right 
panels respectively) for the full-SDE setup.  See text for details of the parameter configurations.}
\end{figure}

\begin{figure}[t]
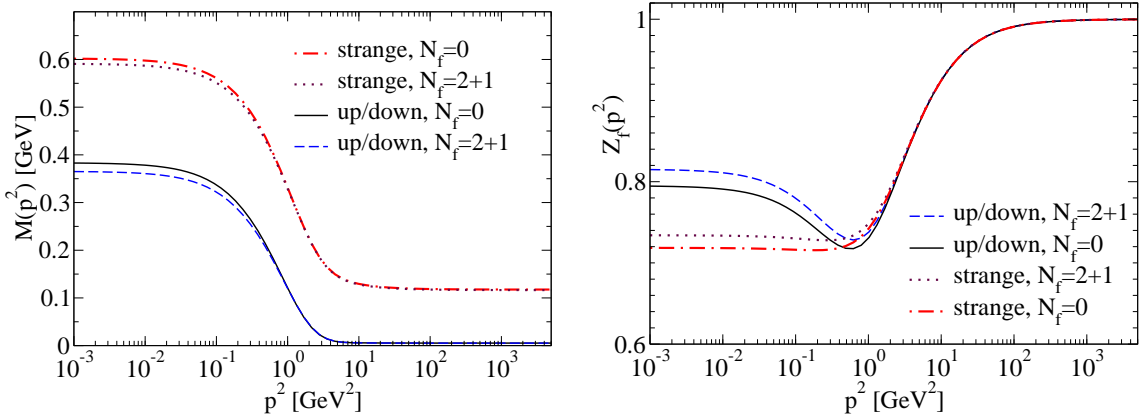

\vspace{1cm}
\mbox{\epsfig{figure=quarkm.eps,width=7.3cm}\hspace{0.5cm}\epsfig{figure=quarkz.eps,width=7.3cm}}\\
\caption{\label{fig:qspace_phen}Comparison of the quenched and unquenched quark mass and wave-functions (left and right 
panels respectively) for the phenomenological model.  See text for details of the parameter configurations.}
\end{figure}

The up/down and strange quark propagator functions for the full-SDE framework and for the phenomenological model are plotted 
in Fig.~\ref{fig:qspace_full} and Fig.~\ref{fig:qspace_phen}, respectively.  Both sets of results for the quarks clearly 
exhibit a large degree of dynamical chiral symmetry breaking in the infrared region and we see the aforementioned effects 
of a reduction of the light quark mass function in the infrared once quark-loops are taken into account.  
For large momenta the numerical solutions of the full-SDE setup reproduce the logarithmic running known from resummed 
perturbation theory (see \cite{Fischer:2003rp} for a detailed discussion of the ultraviolet properties of the quark mass function). 
Clearly, with the effective interaction of the phenomenological 
model being localised in the mid-momentum region, these asymptotic behaviours are not present.  For intermediate momenta for 
all the quark flavors, we observe noticeable unquenching effects for the full-SDE setup which are not present in the 
phenomenological model. In the infrared, the effect of unquenching is that $Z_f(p^2)$ is raised slightly and $M(p^2)$ is 
lowered in all cases.  There is, in addition, a striking difference in the behaviour of the quark wave function $Z_f(p^2)$ 
when comparing the two models (though the unquenching effects are similar to both). In the up-quark of the phenomenological 
model, the localised interaction results in bending $Z_f$ back upwards as one moves into the infrared region.  This effect 
is suppressed for the strange quark.


The last point of note concerns possible singularities in the quark functions in the complex plane.  These can 
readily be identified with our technique for calculating the quark functions in the complex plane (see appendix \ref{app:C}), or 
using Schwinger-function techniques \cite{Alkofer:2003jj}.  We find that the nearest singularities to the origin occur as pairs of 
complex conjugate poles in the timelike region.  For the chiral quarks of the phenomenological model they are found at roughly 
$(-0.26\pm0.17\imath) \mathrm{GeV}^2$ (quenched) and $(-0.23\pm0.13\imath) \mathrm{GeV}^2$ (unquenched), whereas for the 
finite mass quarks they are further out in the timelike region.  With the full-SDE setup the poles are further away in 
the timelike region as compared with the phenomenological model and there is little difference between the quenched and 
unquenched cases.  The likely explanation for the full-SDE poles being further from the origin is again the observation that the 
associated interaction is not localised to a specific region but rather extends throughout the full momentum range.  We conclude 
that the unquenching has a slight effect on the position of the singularities but will not affect the BSE calculations of 
later sections.  Indeed it has been checked during the numerical calculations that no singularities appear within the boundaries of 
the integration region associated with the BSE for any of the results shown in the paper.

\subsection{Numerical Results for Mesons: Physical quark masses, $N_f=2+1$}

In this subsection we investigate two scenarios: quenched and unquenched 
with three physical quark masses, {\it i.e.} two light quarks representing the up/down quark and one heavier quark
representing the strange quark. (Our results for three degenerate quark masses are presented in the next subsection.)

To start, let us explain how we determine physical observables. In the case of the full-SDE setup we have to 
use one experimental quantity (here the pion decay constant $f_\pi=130.7\,\mathrm{MeV}$) to convert the scale generated by
dimensional transmutation to physical units. Furthermore we determine the quark masses $m_{u/d}$ and 
$m_s$ by the pseudoscalar observables $m_{\pi}=139.6\,\mathrm{MeV}$ and $m_K=493.7\,\mathrm{MeV}$ \cite{Eidelman:2004wy}.
To roughly compare with the values of the particle data group we determine $m_{u/d}$ and $m_s$ from the quark mass function at large
momenta and subsequently evolve them down to $\mu=2\,\mathrm{GeV}$. To this end we extract the one-loop scale 
$\Lambda_{\widetilde{MOM}}^{N_f=0}=0.95\,\mbox{GeV}$ and $\Lambda_{\widetilde{MOM}}^{N_f=3}=1.09\,\mbox{GeV}$
from the ultraviolet behaviour of the running coupling. Note that we have employed a $\widetilde{MOM}$ scheme, thus 
these scales have the expected 
magnitude\footnote{For a discussion of the relation of a $\widetilde{MOM}$-scheme to the $\overline{MS}$-scheme see 
section IV of Ref.~\cite{Becirevic:1999hj}. Based on a three loop calculation the authors obtained the 
relation $\Lambda_{\overline{MS}} \simeq 0.346 \Lambda_{\widetilde{MOM}}$. In our case this would result in 
the value $\Lambda_{\overline{MS}}=0.329\, \mbox{GeV}$ for zero flavors.}. In the case of the phenomenological model the 
scale is set by the paramater $d$ in the effective interaction, where in addition we take the fixed value 
$\w=0.5\,\mathrm{GeV}$ and $\mu=1\,\mathrm{GeV}$.  In order not to obscure the present discussion, details of the procedure for 
fitting $\mu$ are given in Appendix~\ref{app:D}. The quark masses are determined by fitting to the experimental pion and kaon masses.

The final values for the parameter sets with the resulting pseudoscalar meson masses, leptonic 
decay constants and vector meson masses are given in Table~\ref{tab:fpar}. When fitted to the experimental
pion and kaon masses the resulting up/down and strange-quark masses are lowered when quark loop effects are
taken into account. This has also been observed in corresponding lattice simulations \cite{CP-PACS,JLQCD}.
We furthermore see that the results for $f_K$ and $m_{\ro}$ are quite insensitive to whether or not the system is unquenched.  
This leads to the conclusion that once the interaction has been fitted to the observable pseudoscalar observables, 
the vector meson mass is largely fixed.  The likely explanation for this is that the ground state pseudoscalar and 
vector mesons are both states with the lowest orbital angular momentum ($L=0$) in the sense of the naive (quantum 
mechanical) quark model -- meaning that they are determined largely by the lowest spin contributions of the BS 
kernel given by the ladder approximation.  The interaction then plays the same role in both channels, hence the 
similarity in results.  Note that the $\rho$ meson calculated within the framework of the truncated Bethe-Salpeter equation here 
refers to a pure quark-antiquark meson with no allowed decay channel - the physical $\rho$ meson has a non-trivial decay width.  
Thus, one must allow for some change in the calculated mass if one were to take the decay process into account.  However, our 
results for $m_{\rho}$, especially in the full-SDE setup do seem low, although not unreasonable.

\begin{table}[t]
\caption{\label{tab:fpar}Parameter sets and results for $m_{\pi}$, $f_{\pi}$, $m_K$ $f_K$ and $m_{\ro}$ for the quenched
case ($N_f=0$), the unquenched case with three degenerate 'sea'-quarks ($N_f=3$) and the physical quark 
configuration case ($N_f=2+1$) with two up/down quarks and one strange quark.  In the phenomenological model $\w=0.5\,\mathrm{GeV}$ 
is held fixed and $\mu=1\,\mathrm{GeV}$. In the full-SDE setup we give values for the quark masses read off at a large scale and 
evolved down to $\mu = 2\,\mathrm{GeV}$ according to their one-loop behaviour.}
\begin{center}\begin{tabular}{|c||c|c|c||c|c|c|c||c|}\hline
&$d$($\mathrm{GeV}^{-2}$)&$m_u$($\mathrm{MeV}$)&$m_s$($\mathrm{MeV}$)&
$m_{\pi}$($\mathrm{MeV}$)&$f_{\pi}$($\mathrm{MeV}$)&$m_K$($\mathrm{MeV}$)&$f_K$($\mathrm{MeV}$)&$m_{\ro}$($\mathrm{MeV}$)
\\\hline\hline
model:    &      &      &       &      &      &      &      &     \\\hline
$N_f=0$   &14.92 &5.26 &117.5   &139.6 &130.7 &493.7 &152.4 &737.6\\\hline
$N_f=3$   &16.94 &5.17 &        &139.6 &130.7 &      &      &734.7\\\hline
$N_f=2+1$ &16.66 &5.18 &116.4   &139.6 &130.7 &493.7 &153.0 &735.0\\\hline\hline
&$\Lambda_{MOM}$($\mathrm{GeV}$)&$m_u^{2 GeV}$($\mathrm{MeV})$&$m_s^{2 GeV}$($\mathrm{MeV}$)&
$m_{\pi}$($\mathrm{MeV}$)&$f_{\pi}$($\mathrm{MeV}$)&$m_K$($\mathrm{MeV}$)&$f_K$($\mathrm{MeV}$)&$m_{\ro}$($\mathrm{MeV}$)
\\\hline\hline
full-SDE: &      &      &       &      &      &      &      &     \\\hline
$N_f=0$   &0.95  &4.17  &88.2   &139.7 &130.9 &494.5 &165.6 &708.0\\\hline
$N_f=3  $ &1.16  &4.06  &       &139.7 &130.8 &      &      &690.0\\\hline
$N_f=2+1$ &1.09  &4.06  &86.0   &140.0 &131.1 &493.3 &169.5 &695.2\\\hline\hline
PDG\cite{Eidelman:2004wy}&&&    &139.6 &130.7 &493.7 &160.0 &770  \\\hline
\end{tabular}\end{center}\end{table}

\subsection{Numerical Results for Mesons: Degenerate quark masses, $N_f=3$}

Let us now consider the degenerate quark configuration case, with $N_f=3$ identical quarks.  To fit the parameters, we use a 
similar approach as that for the physical case, fitting to the pseudoscalar sector, though only the pion observables are 
relevant since there are no kaons. The parameters used 
and the resultant pion observables, along with the $\rho$ meson mass are shown in Table~\ref{tab:fpar}.  We see that the 
unquenched results for $m_{\ro}$ are only slightly changed by the different unquenching configuration. 

\begin{figure}[t]
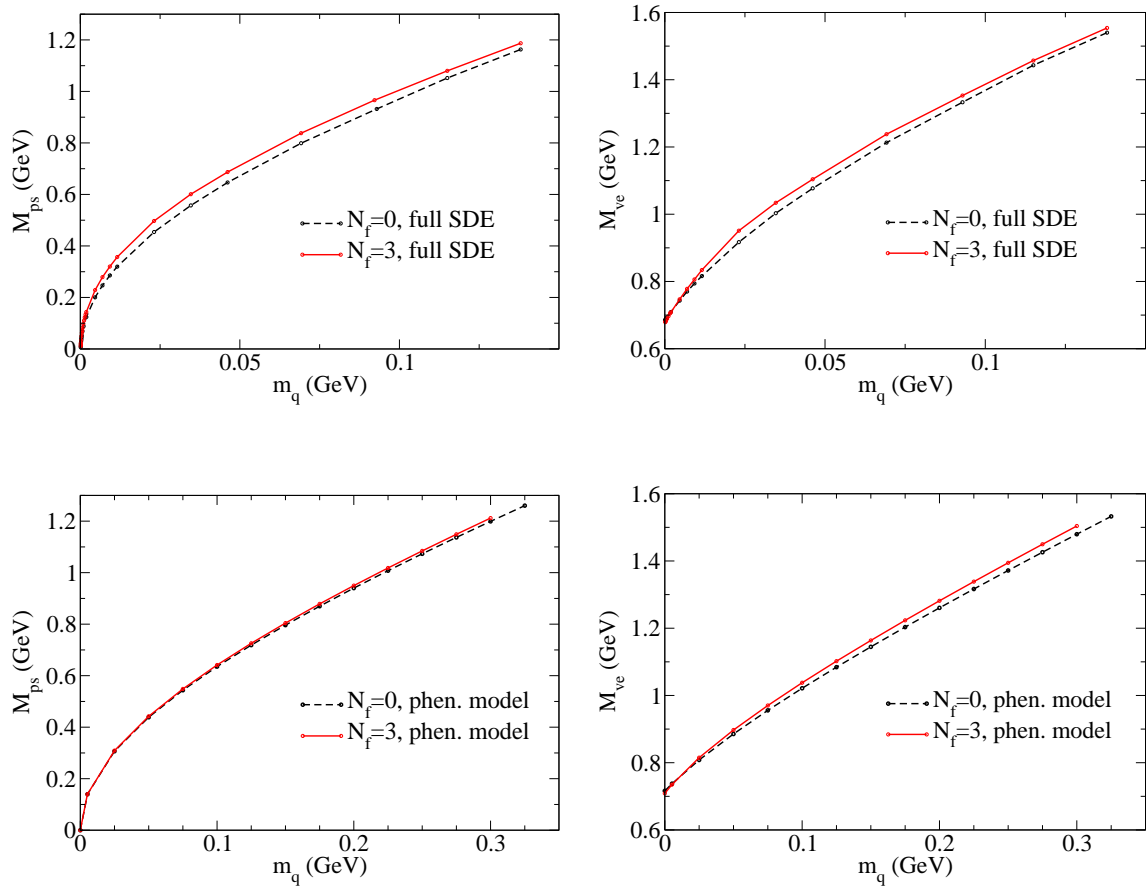

\vspace{1cm}
\mbox{\epsfig{figure=pi_full.eps,width=7.3cm}\hspace{0.5cm}\epsfig{figure=rho_full.eps,width=7.3cm}}\\
\vspace{1cm}
\mbox{\epsfig{figure=pi.eps,width=7.3cm}\hspace{0.5cm}\epsfig{figure=rho.eps,width=7.3cm}}\\
\caption{\label{fig:plot1}Pseudoscalar (left panels) and vector (right panels) meson masses as functions of 
the quark mass parameter.  We compare the quenched and degenerate unquenched cases for both the full-SDE 
(upper panels) and phenomenological model (lower panels).  Recall that there is no direct comparison between 
the quark masses of the full-SDE and phenomenological model.}
\end{figure}

To further study the effects of unquenching we take the quenched and degenerate unquenched parameter configurations and plot 
the pseudoscalar and vector meson masses as a function of the quark mass parameter, keeping all other parameters 
fixed\footnote{The unquenched case with physical 'sea'-quark configurations is not considered because with the heavy 
meson masses generated by the 'valence'-quarks, the light quark propagators of the interaction need to be evaluated in 
a domain which includes the propagator singularities.  This introduces technical problems that are beyond the scope of 
this work.}. The results are shown in Fig.~\ref{fig:plot1}. We see that for the full-SDE setup, both the pseudoscalar 
and vector masses with larger quark masses are increased $\sim 30\mathrm{MeV}$ when we include the unquenching effects.  
For the phenomenological model we see the same trend though the effect is less pronounced.  Notice that as $m_q\rightarrow 0$ 
(the chiral limit) in both models, the pseudoscalar meson mass vanishes in accordance with the chiral symmetry considerations.  
This is a good check that the AXWTI has been faithfully maintained throughout the numerical procedure.

\begin{figure}[t]
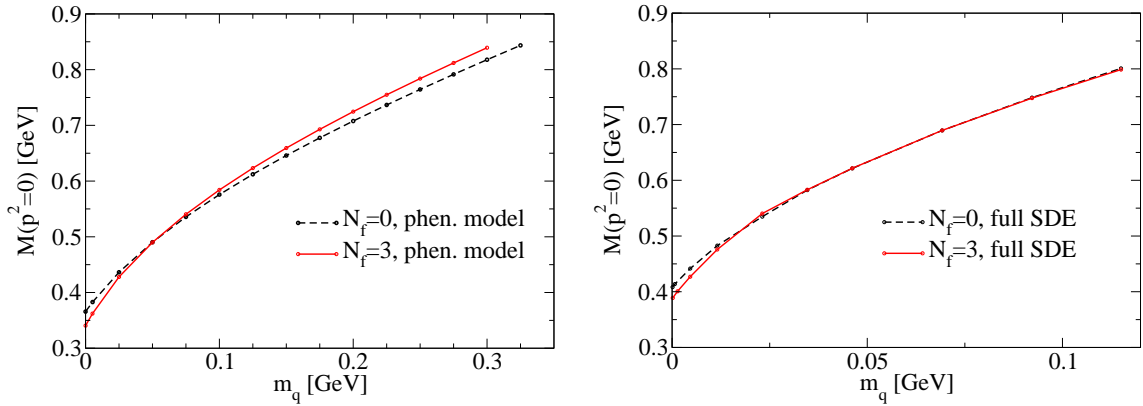

\vspace{1cm}
\mbox{\epsfig{figure=mass.eps,width=7.3cm}\hspace{0.5cm}\epsfig{figure=mass_full.eps,width=7.3cm}}\\
\caption{\label{fig:plot2}Quark mass function at $p^2=0$ as a function of the quark mass parameter.  We 
compare the quenched and degenerate unquenched cases.}
\end{figure}

\begin{figure}[t]
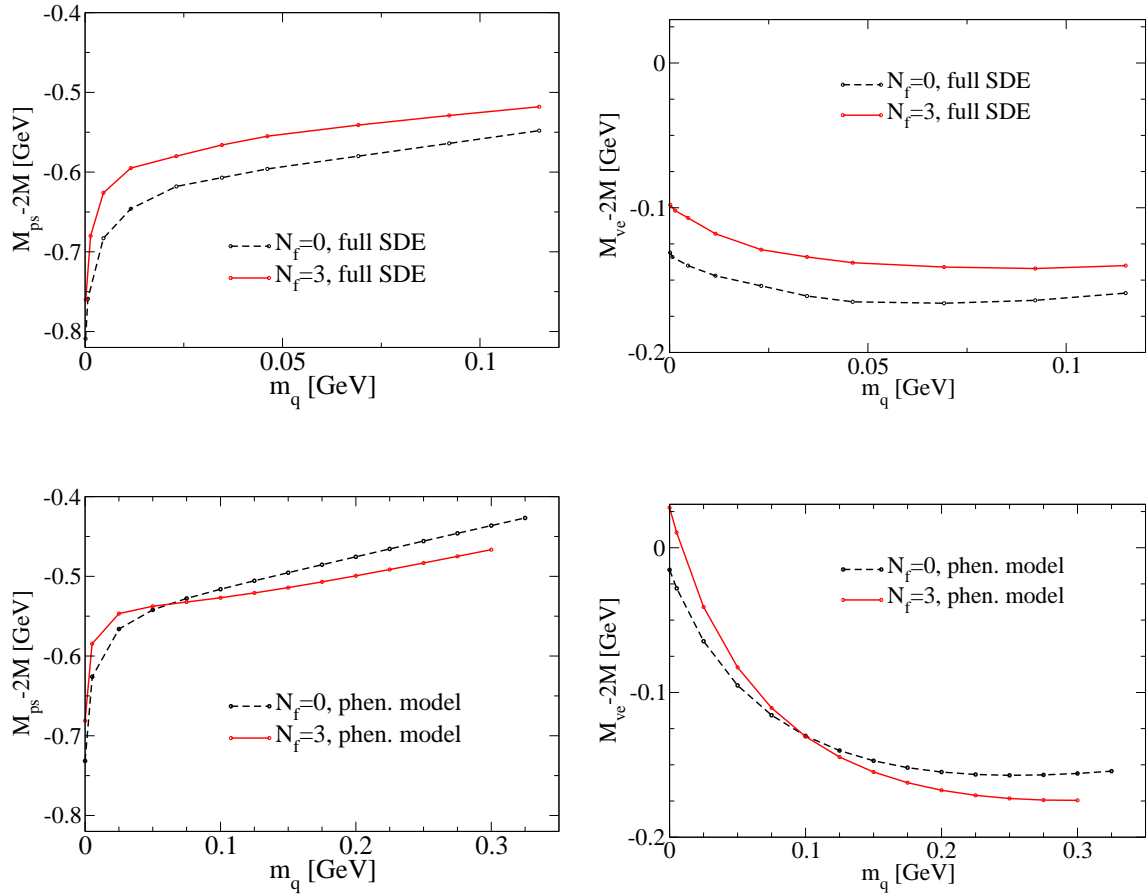

\vspace{1cm}
\mbox{\epsfig{figure=bps_full.eps,width=7.3cm}\hspace{0.5cm}\epsfig{figure=bve_full.eps,width=7.3cm}}\\
\vspace{1cm}
\mbox{\epsfig{figure=bps.eps,width=7.3cm}\hspace{0.5cm}\epsfig{figure=bve.eps,width=7.3cm}}\\
\caption{\label{fig:plot3}The difference between the pseudoscalar (left panel) and vector (right panel) 
meson masses and twice the quark mass function $M(p^2=0)$ as functions of the quark mass parameter.  We 
compare the quenched and degenerate unquenched cases.}
\end{figure}

Given the differences in the models, one is led to investigate in more detail the relationship between the meson masses, the 
interaction and the quark mass functions and how this relationship is altered when the quark loop is included.  We plot the 
quark mass function at $p^2=0$ as a function of the quark mass parameter in \fig{fig:plot2}.  This serves to give a measure 
on the effect of unquenching for finite mass parameter quarks. We also plot in \fig{fig:plot3} the difference between the 
meson masses and twice the quark mass function $M(p^2=0)$.  In the static, quantum mechanical model of mesons this would be 
interpreted as a binding energy due to the interaction.  We point out, however, that in the (relativistic) Bethe-Salpeter approach 
here this interpretation does not hold.  We shall refer to this difference as the 'binding' but the reader should bear in mind 
that this is a loose definition of the term.  We see that in \fig{fig:plot3} the unquenching does indeed have a significant 
effect on the internal behaviour of the BSE.

Expressed in the form of figs.~\ref{fig:plot2} and \ref{fig:plot3} one can now clearly see the difference in the two models.  
In the full-SDE setup at finite quark masses the quark mass function becomes insensitive to unquenching.  This can be 
intuitively explained because the quark loops become suppressed by their mass.  However, the overall meson masses show a 
$\sim30\mathrm{MeV}$ increase, reflected in the binding.  Thus in the full-SDE setup, the unquenching of the interaction 
has an effect in the BSE but not in the quark mass function.

In distinction, for the phenomenological model the situation is more complex.  At low quark masses the quark mass functions 
and the binding do show the same behaviour as for the full-SDE setup - but of course for low quark masses all mesons are in 
the vicinity of the fixed physical point from which the parameters are fitted.  As one increases the quark mass though, the 
curves for the quenched and unquenched quark mass functions and binding cross.  Looking at the parameter sets, one sees that 
in the unquenched case the overall magnitude $d$ is larger.  As the quark masses are increased, the quark loop becomes 
suppressed and one is left with a stronger interaction, hence the increased quark mass function. (Recall that in the full-SDE 
setup there is no fitted parameter corresponding to the overall magnitude, hence the different behaviour of the quark mass 
function).  Since for small $m_q$ the unquenched mass function is lower than the quenched case eventually the curves cross.  
However, the meson mass does not increase so significantly under unquenching for larger $m_q$, even though both the interaction 
and the quark mass function do increase.  The resolution to this apparent dichotomy lies in the fact that for larger quark mass 
parameters, the relative contribution of the bare quark mass to the total meson mass is greater and the effect of increasing 
the interaction and quark mass function is damped (recall that the quark mass function referred to here is $M(p^2=0)$).

This discussion of the detailed effects of unquenching serves to highlight that we are not dealing with a system where the 
total mass is given by the sum of the constituent masses and their binding energy but rather with a relativistic system 
whose constituents all depend non-trivially upon each other and crucially, all the components have varying effects at different 
momentum scales.  The two models exhibit different behaviour when quark loops are included and this is because of the 
different forms of the interaction.  The overall properties of the meson masses are largely determined by the 
integrated strength of the interaction. However, the closer inspection of unquenching for finite quark masses does indeed 
reveal that the form of the interaction plays a role.

\begin{figure}[t]
\vspace{1cm}
\mbox{\epsfig{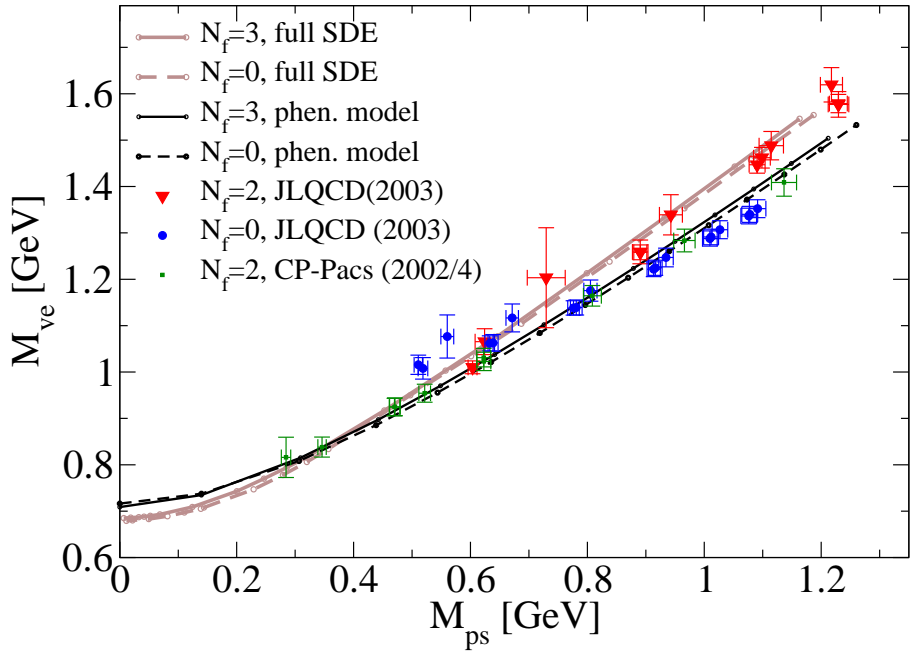}}\\
\caption{\label{fig:plot4}Vector meson masses as a function of pseudoscalar meson masses.  We compare the 
quenched and degenerate unquenched cases. The lattice results are taken from refs.~\cite{CP-PACS,JLQCD}.}
\end{figure}

Finally, we make contact again to corresponding lattice calculations \cite{CP-PACS,JLQCD}. Since it is most conclusive to 
compare 'physical' quantities we plot the vector meson masses vs. the corresponding pseudoscalar meson masses. 
For large quark masses one may expect that unquenching effects in these quantities will be mainly determined by the 
interaction alone, since the decay of the vector meson to two pseudoscalar mesons is no longer kinematically allowed 
(though in principle unquenching effects involving vertex corrections beyond the truncation schemes discussed here may 
also play a role). The comparison to the lattice is therefore a genuine test of our two setups. 
From the plot in \fig{fig:plot4} we find good agreement between the curves from both of our setups and the
available lattice data. It is apparent that the effects due to unquenching are smaller than both the differences 
due to the different truncation schemes and the systematic errors of the lattice results.
Because we are plotting the more physical meson masses and not the (incomparable) quark mass parameters, the difference 
between the quenched and unquenched curves for both of our setups becomes the same -- the curve for the unquenched vector 
meson mass as a function of the pseudoscalar meson mass is slightly raised for large values of the pseudoscalar meson mass.  
For small masses we observe a nonlinear relation of the two masses reflecting the different roles of the
vector and the pseudoscalar meson sector in chiral limit.

\section{Summary and Conclusions}\label{sec:Conclusions}

We have investigated the effects of unquenching the gluon polarisation when applied to the \BS framework for light 
pseudoscalar and vector meson masses. To obtain model-independent information on these effects we considered two 
distinct methods of calculating the corresponding gluon and quark propagators self-consistently.  The first uses 
a sophisticated truncation of the \DS equations for the ghost, gluon and quark propagator. The second uses a 
(simpler) phenomenologically motivated form for the quenched Yang-Mills sector of the theory. 
In both approaches, we explicitly took quark loop contributions to the gluon polarisation into account. 
Employing the resultant propagator functions in conjunction with a truncation for 
the \BS equation we calculated light pseudoscalar and vector meson masses.  The forms of the truncated \BS kernel are 
such that the axialvector Ward-Takahashi identity for color singlet, flavor non-singlet channels is obeyed, thereby 
ensuring that the generic features of the chiral limit of the pseudoscalar meson sector are manifest.
Though the resulting effective quark-gluon interactions in our two approaches are rather different, both produce
similar qualitative effects in meson observables. This finding enforces our general belief that the lowest meson
states are dominated by the pattern of dynamical chiral symmetry breaking and are not largely affected by the 
detailed structure of the interaction.

We observed sizeable unquenching effects due to the inclusion of quark loops to the gluon polarisation at various 
stages of our investigation. For the ghost and gluon propagator functions, calculated in the full \DS truncation scheme, 
these effects occur predominantly in the intermediate momentum region. At momenta around  $1\mathrm{GeV}$ the screening of 
quark-antiquark pairs moderates the bump in the gluon propagator considerably, whereas the infrared behaviour of both 
propagators is unaffected. For the quark propagator functions we found slight effects also in the infrared: the 
quark mass and wave functions for up/down and strange quark masses are reduced by about 10\% when unquenched. 
These results agree with previous findings \cite{Fischer:2003rp} and have recently been confirmed in lattice 
simulations \cite{Bowman:2004jm,Bowman:2005vx}. We determined the location of the singularities of the quark propagator 
in the complex momentum plane and found a pair of complex conjugated poles, which are hardly affected by the inclusion of 
quark loop effects. Furthermore we found only very small changes in the chiral condensate as long as the number of 
flavors $N_f \le 3$, indicating a large critical number of flavors for the chiral phase transition. All these effects 
are model independent in the sense that we observed them in the full SDE-setup as well as in our phenomenological 
model (where assessible).

The main focus of the present work has been on the spectrum and decay constants of pseudoscalar and vector mesons. 
Here, the physical pion and kaon masses along with the pion leptonic decay constant have been used to determine the scale of the
interaction as well as the up/down and strange quark masses. The resulting masses at the physical point are smaller when 
quark loop effects are taken into account. Again this agrees with observations in corresponding lattice 
simulations \cite{CP-PACS,JLQCD}. The kaon leptonic decay constant and vector meson mass were in rough agreement with 
available data, whereas the vector meson mass turned out slightly too low. The vector mass is subject to 
corrections due to a finite rho decay width \cite{Watson:2004jq} and further structure in the quark-gluon 
vertex \cite{Bhagwat:2004hn,Watson:2004kd}, which both were beyond the scope of the present work. Importantly however, 
our results for $f_K$ and $m_\rho$ were not significantly altered when quark loops were included.

To further discuss the role of the quark loop contribution to the gluon polarisation, we compared quenched and unquenched 
mesons with degenerate sea-quark configurations. Keeping the scale of the interaction fixed, we studied the quark mass 
function as well as the pseudoscalar and vector meson masses as functions of the quark mass parameter. As expected, 
both of our approaches preserve the Goldstone character of the pion in the chiral limit. For small quark masses the 
observables depend nonlinearly on the quark mass parameter. For large quark masses the two schemes
display different behaviour under unquenching.  The full \DS truncation results exhibit an increase in the meson masses 
when quark loops are included, whereas for the phenomenological model this increase is much smaller.  

Furthermore we studied the vector meson masses as a function of the pseudoscalar meson mass and compared with recent 
lattice results from various groups.  The resultant curves for the two schemes, though different from each other, are nonetheless 
in general agreement with the lattice data. For pion masses below 240 MeV, where no lattice data are available, our results
show a nonlinear dependence of the vector meson mass on the pseudoscalar one.
The effect of unquenching -- when viewed as a function of the pseudoscalar meson mass -- becomes the same for both schemes: 
the vector meson mass is slightly increased when quark loops are taken into account. This trend is also seen in the 
lattice simulations \cite{CP-PACS,JLQCD}, where the effect is even more pronounced.  However, these unquenching effects are small 
compared to the differences between both the truncation schemes we employed and the systematic errors of the lattice results.

\acknowledgments

The authors thank M.~R.~Pennington for useful discussion.  C.~S.~Fischer would also like to thank 
A.~Krassnigg for further discussions.  This work has been supported by the Deutsche For\-schungsgemeinschaft
(DFG) under contract Fi 970/2-1 and the Virtual Institute for Dense Hadronic Matter and QCD Phase Transitions.

\appendix
\section{\label{app:A} Conventions for Renormalisation in the full SDE Truncation Scheme}

In the full \DS equation truncation, renormalised dressing functions are denoted by an overbar.  They are related to their 
unrenormalised counterpart via the renormalisation constant $Z_i(\mu,\La)$ where $\mu$ is the renormalisation scale and $\La$ 
is the regularisation scale.  The multiplicative renormalisation constants are defined as follows:
\bea
S(p,\La)&=&Z_2(\mu,\La)\ov{S}(p,\mu)~~~,\\
D(p,\La)&=&Z_3(\mu,\La)\ov{D}(p,\mu)~~~,\\
G(p,\La)&=&\tilde{Z}_3(\mu,\La)\ov{G}(p,\mu)~~~,\\
\G_{\ro}(p_i,\La)&=&Z_{1F}^{-1}(\mu,\La)\ov{\G}_{\ro}(p_i,\mu)~~~,\\
\tilde{\G}_{\ro}(p_i,\La)&=&\tilde{Z}_1^{-1}(\mu,\La)\ov{\tilde{\G}}_{\ro}(p_i,\mu)~~~,\\
\G_{\ro\nu\la}(p_i,\La)&=&Z_1^{-1}(\mu,\La)\ov{\G}_{\ro\nu\la}(p_i,\mu)~~~.
\eea
The renormalised fermion mass function is obtained with the relation $Z_2M_{\La}=Z_4\ov{M}_p$ and the 
running quark mass $\ov{M}_p$ is independent of the renormalisation scale $\mu$.  The coupling is renormalised 
according to
\be
g^2(\La)=\frac{\tilde{Z}_1^2}{Z_3\tilde{Z}_3^2}\ov{g}^2(\mu)
\ee
and the Slavnov-Taylor identity, which ensures the universality of the renormalised coupling, is expressed as
\be
\frac{Z_3}{Z_1}=\frac{\tilde{Z}_3}{\tilde{Z}_1}=\frac{Z_2}{Z_{1F}}~~~.
\ee


\section{\label{app:B} A Truncation of the ghost and gluon SDEs}

\begin{figure}[t]
\centerline{\epsfig{file=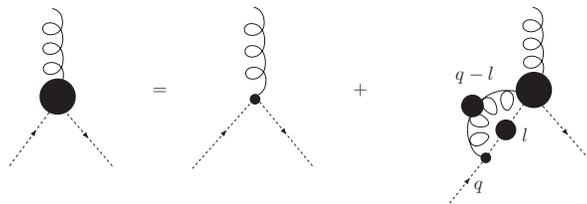,width=80mm}}
\caption{Ghost-gluon vertex DS equation.}
\label{SDE-ghg}
\end{figure}

Here we shortly summarise the truncation for the ghost- and gluon-SDEs as introduced in 
refs.~\cite{Fischer:2002hn,Fischer:2003rp,Fischer:2003zc}. In order to close the Eqs.~(\ref{ghost-SDE}) and
(\ref{gluon-SDE}) we have to specify expressions for the ghost-gluon vertex, the three-gluon vertex and
the four-gluon vertex. Our ansatz for the quark-gluon vertex has been discussed in the main body of this work,
cf. Eq.(\ref{quark-gluon-vertex}).

The crucial property of the ghost-gluon vertex\footnote{Note that the color structure of the ghost and 
gluon propagators is just a trivial delta-function. Thus the color parts of all dressed Yang-Mills 
vertices get directly contracted with the antisymmetric structure constant $f^{abc}$ of their bare 
counterparts in the respective loops of the ghost and gluon SDEs. Possible color-symmetric 
contributions to the fully dressed vertices therefore do not contribute in these equations. We
therefore use only the antisymmetric structure constants from the start.}
\be
\Gamma_\mu^{abc}(p,q) = \Gamma_\mu(p,q) f^{abc}
\ee
is that both, incoming and outgoing ghost momentum factorise in Landau gauge \cite{Taylor:ff}. Indeed, factorisation 
of the outgoing ghost momentum $p$ is immediate and occurs in all linear covariant gauges.  Factorisation
of the incoming ghost momentum $q$ can be seen easily from its \DS equation, Fig.~\ref{SDE-ghg}, and the 
transversality of the gluon propagator $D_{\mu \nu}$, which implies that $l_\mu D_{\mu \nu}(l-q) = q_\mu D_{\mu \nu}(l-q)$.
Thus, provided the gluon-ghost kernel in Fig.~\ref{SDE-ghg} is well behaved in the infrared (i.e. not too singular),
we obtain a bare vertex for $q_\mu \rightarrow 0$. Indeed at least one self-consistent infrared solution of the whole
tower of SDE exists \cite{Alkofer:2004it}, in which the gluon-ghost kernel indeed is singular, but not singular enough 
to spoil the argument. If this is true, then a bare ghost-gluon vertex 
\be
\Gamma_\mu (q,p) = iq_\mu 
\label{bare-gg-vertex}
\ee
provides a good approximation to the full vertex in the infrared and also in the ultraviolet region 
of momentum. Indeed, recent lattice calculations \cite{Cucchieri:2004sq} and also a study in the 
SDE-approach \cite{Schleifenbaum:2004id} support this approximation.

One knows from the self-consistent infrared solution \cite{Alkofer:2004it}, that the ghost-loop is the leading
diagram of the gluon-SDE in the infrared. Although the three-gluon and four-gluon vertices are singular in this
scenario, they are not singular enough to overcome the suppression induced by the gluon dressing functions. Thus effects
from the four-gluon vertex on the gluon-propagator are negligible in the infrared. As they are also subleading in the 
ultraviolet and moreover are technically complicated to evaluate we neglect these effects from the start by choosing
the four-gluon vertex to be zero.

On the other hand, although the effects from the gluon-loop including the three-gluon vertex are also negligible
in the infrared, they contribute to the one-loop running of the propagator in the ultraviolet and have considerable
impact on the mid-momentum behaviour of the propagator. One therefore has to include a suitable ansatz for the three-gluon
vertex. Such an ansatz has been constructed and investigated in refs.~\cite{Fischer:2002hn,Fischer:2003rp,Fischer:2003zc} and 
is given by
\begin{equation}
\Gamma_{\mu \nu \lambda}^{abc}(k,p,q) = \Gamma_{\mu \nu \lambda}^{abc,0}(k,p,q) H_{3g}(k,p,q)
\end{equation}
where $\Gamma_{\mu \nu \lambda}^{abc,0}(k,p,q)$ is the bare vertex. The 
dressing function $H_{3g}$ is given by 
\begin{equation} \label{3g}
H_{3g}(k,p,q)=\frac{1}{Z_1} \frac{G(q)^{(-2-6\delta)}}{Z(q)^{(1+3 \delta)}} 
\frac{G(p)^{(-2-6 \delta)}}{Z(p)^{(1+3 \delta)}}\; ,
\end{equation} 
with the momenta $q$ and $p$ running in the loop. Here the anomalous dimension 
of the ghost propagator, $\delta= -9N_c/(44N_c-8N_f)$, together with the vertex renormalization 
constant $Z_1$ ensure the correct ultraviolet running of the vertex with momenta and 
renormalization scale. Furthermore this choice leads to cutoff independent ghost and 
gluon dressing functions in the continuum.


\section{\label{app:C} Solving the SDE in the Complex Plane}

In the (Euclidean space version of the) Bethe-Salpeter equation, one is looking for a solution $P^2=-M^2$ which necessarily 
introduces complex momenta into the problem.  The quark propagator must be evaluated at momenta within a parabola 
$k^2-M^2/4\pm\imath\sqrt{k^2M^2}$ in the complex plane.  In order to achieve this, we must solve the coupled quark and 
gluon propagator SDEs in the complex plane and we shall do this using a combination of contour integrals and UV asymptotic 
expansion.  Expanding the integral measure and paying attention to the flavor index $i$, the coupled equations 
(\ref{eq:a1},\ref{eq:b1},\ref{eq:s1}) become
\bea
a^i(x)&=&1+\frac{4}{3(2\pi)^3}\int_0^{\infty}dy\,y\int_{-1}^1dz\sqrt{1-z^2}\,g^2\De(y)v^i(x+y-2\sqrt{xy}z)
        \left[1-3{\frac{\sqrt{y}}{\sqrt{x}}}z+2z^2\right],\label{eq:a3}\\
b^i(x)&=&m^i+\frac{4}{3(2\pi)^3}\int_0^{\infty}dy\,y\int_{-1}^1dz\sqrt{1-z^2}\,g^2\De(y)s^i(x+y-2\sqrt{xy}z)
         \left[3\right],\label{eq:b3}\\
\Si(x)&=&\sum_{i=u,d,s}\frac{4}{3(2\pi)^3}\int_0^{\infty}dy\,y\int_{-1}^1dz\sqrt{1-z^2}v^i(y)v^i(x+y-2\sqrt{xy}z)
         \left[\frac{y}{x}(1-4z^2)+3\frac{\sqrt{y}}{\sqrt{x}}z\right],\label{eq:s3}
\eea
where $x=p^2$, $y=k^2$ and $z=p\s k/\sqrt{p^2k^2}$.  Alternatively, one can choose a symmetric momentum routing.
Equations (\ref{eq:a3},\ref{eq:b3},\ref{eq:s3}) then become
\bea
a^i(x)&=&1+\frac{4}{3(2\pi)^3}\int_0^{\infty}dy\,y\int_{-1}^1dz\sqrt{1-z^2}\,g^2\De(y_-)v^i(y_+)\left[-\frac{1}{2}
         +3{\frac{\sqrt{y}}{\sqrt{x}}}z+2\frac{x/4+yz^2-\sqrt{xy}z}{x/4+y-\sqrt{xy}z}\right],\label{eq:a4}\\
b^i(x)&=&m^i+\frac{4}{3(2\pi)^3}\int_0^{\infty}dy\,y\int_{-1}^1dz\sqrt{1-z^2}\,g^2\De(y_-)s^i(y_+)\left[3\right],\label{eq:b4}\\
\Si(x)&=&\sum_{i=u,d,s}\frac{4}{3(2\pi)^3}\int_0^{\infty}dy\,y\int_{-1}^1dz\sqrt{1-z^2}v^i(y_-)v^i(y_+)
         \left[\frac{y}{x}(1-4z^2)+\frac{3}{4}\right],\label{eq:s4}
\eea
where $y_{\pm}=x/4+y\pm\sqrt{xy}z$.  The reason for the two momentum routings is the following:  At large 
values of the radial integral variable $y$, the asymmetric momentum routing has factors of either $g^2\De(y)$ or $v^i(y)$ 
which vanish quickly for large $y$.  The fidelity of the angular integration is thus less important even when $x$ and 
$y$ are both large (large $x$ here refers to spacelike momenta).  In the symmetric case at large $x$ and $y$, the 
angular integration involves integrating over functions which vary considerably over the entire momentum range and so 
is numerically much more difficult.  For this reason, although the two sets of equations are formally equivalent 
(ignoring for now the possible complications of the UV-cutoff), numerically we use the asymmetric routing at large 
$x$ such that the angular integral can be more readily done.  The use of the symmetric routing comes about in the 
context of the contour integrals below.

\begin{figure}[t]
\mbox{\epsfig{figure=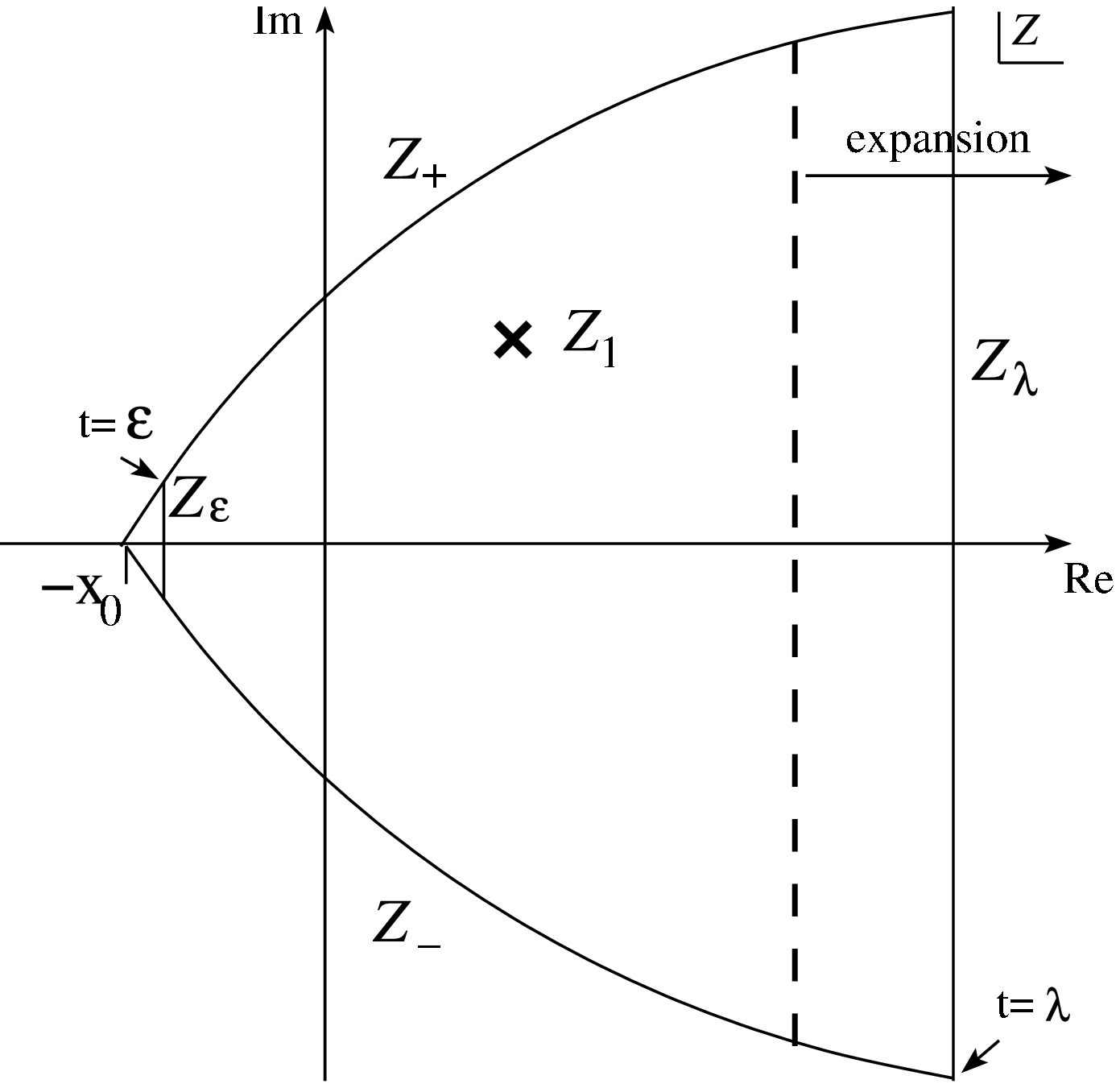,width=10.0cm}}
\caption{\label{fig:parab}Sketch of the closed contour in the complex $z$ plane.}
\end{figure}

Let us now describe how the equations are solved in the complex plane.  At large $\Re{x}$ the functions take asymptotic 
forms
\bea
a(x)&\stackrel{x\rightarrow\infty}{=}&1+\sum_{k=1}^N\frac{a_k}{x^k},\\
b(x)&\stackrel{x\rightarrow\infty}{=}&m+\sum_{k=1}^N\frac{b_k}{x^k},\\
\Si(x)&\stackrel{x\rightarrow\infty}{=}&\si_0\log{x}+\sum_{k=1}^N\frac{\si_k}{x^{k-1}}.
\eea
For a given set of coefficients one can extrapolate these functions at arbitrary $x$ to the complex region, as long as 
$\Re{x}$ is large.  Now consider the contour defined by the lines (shown in \fig{fig:parab})
\bea
z_-(t)=&-x_0+t-2\imath\sqrt{x_0t},& t=[\e,\la],\nonumber\\
z_{\la}(r)=&-x_0+\la+2\imath\sqrt{x_0\la}r,& r=[-1,1],\nonumber\\
z_+(t)=&-x_0+t+2\imath\sqrt{x_0t},& t=[\la,\e],\nonumber\\
z_{\e}(r)=&-x_0+\e-2\imath\sqrt{x_0\e}r,& r=[-1,1].
\eea
With these definitions the integral measure along the closed contour is:
\be
\oint_cdz\rightarrow\int_{\e}^{\la}dt(1-\imath\sqrt{x_0/t})+2\imath\sqrt{x_0\la}\int_{-1}^1dr-\int_{\e}^{\la}
dt(1+\imath\sqrt{x_0/t})-2\imath\sqrt{x_0\e}\int_{-1}^1dr.\label{eq:cgrid}
\ee
For $z_1$ within the contour the Cauchy integral formula holds:
\be
2\pi\imath f(z_1)=\oint_cdz\frac{f(z)}{z-z_1},
\ee
so given the function on the contour one can derive its value at some arbitrary interior point.  Notice that 
numerically this is easy to implement as long as $z_1$ is not close to the contour itself.

Turning to the above coupled system of SDEs we supply a set of starting functions $a^i$, $b^i$ and $\Si$ along 
the contour and along the spacelike axis, from which one can obtain the asymptotic expansion coefficients.  From 
this we can derive the functions $v^i$, $s^i$ and $g^2\De$ using the expansion (for large real parts of the argument) 
or the contour integration (for small arguments) at those values needed to form the integrands of the coupled 
equations.  With this we can then recalculate the functions on the left-hand side of the equations for external $x$ 
again around the contour or on the spacelike axis.  This gives us an iterative procedure to solve the system of 
equations.  As explained above, we use the asymmetric momentum routing for large $\Re{x}$.  However, for small 
$\Re{x}$ we use the symmetric routing since for some point $x$ on the contour we now need the integrand functions 
at points within the smaller parabola associated with $x/4$ and hence well inside the contour.  This ensures that 
the Cauchy integral formula can be reliably implemented.  It also means that if $v^i$ and $s^i$ are singular in the 
complex region, we can spot this by choosing the contour such that the singularity lies between the contour of external 
$x$ and the smaller parabolic domain accessed by the integrands.

We point out that within the iterative procedure there is no interpolation needed, except for the asymptotic region 
where the functions are well known and slowly varying.  Certainly an interpolating scheme could not, even in principle, 
be used in regions where there are singularities present.  In addition, the contour integrals are quick, simple to 
set up and are more accurate than an interpolation.  The last point is the usefulness of the method in solving the 
BSE -- given the converged solution to the SDEs along the spacelike axis and around the contour, one can then obtain 
the quark functions at any point needed within the parabolic BSE domain directly.

The various parameters associated with the numerical implementation are as follows.  The tip of the
 contour is given by the parameter $\e=0.1\mathrm{GeV}^2$ (avoiding the integrable square-root singularity to make 
 the numerics easier), whereas the end is given by $\la=11\mathrm{GeV}^2$.  $\Re{x}\geq10\mathrm{GeV}^2$ is 'large' 
 and we use the asymptotic expansion.  There a numerical limit to how large the overall contour parameter, $x_0$, 
 which defines the size of the complex region that the system accesses can be and this comes from the requirement 
 that the contour integral be reliable and reasonably fast.  In practice it is found that $x_0\leq0.3\mathrm{GeV}^2$ 
 works well.  If quark propagators are needed outside this domain then one can use the SDE with symmetric routing to 
 evaluate the functions, given the solution within.  This gives an upper limit to the domain of applicability 
 of the technique but this is due to the numerical implementation, not the technique itself.

\section{\label{app:D} Discussion of the renormalisation point dependence of the phenomenological model}

In the phenomenological model, the renormalisation scale $\mu$ is a free parameter to be fitted.  Since the model is not 
fully multiplicatively renormalisable, the results do in general depend on $\mu$. We show here how $\mu$ is constrained.  
Let us begin by considering the behaviour of the chiral condensate defined by \eq{eq:cond} and additionally an approximation 
to the pion leptonic decay constant in the chiral limit \cite{Watson:2004kd}
\be
\left(f_{\pi}^0\right)^2=\frac{3}{2\pi^2}\int_0^{\infty}dy\frac{yb_y}{(ya_y^2+b_y^2)^2}
\left[a_yb_y-\frac{y}{2}(a_yb_y'-b_ya_y')\right].
\ee
It is known that this approximation underestimates $f_{\pi}$ by around $10\%$.  We calculate $f_{\pi}^0$ for a range of 
$\mu$ whilst using $<\ov{q}q>^0$ to fit $d$ with $\w=0.5\mathrm{GeV}$ fixed in order to constrain the range of $\mu$.  
The results are shown in Table~\ref{tab:par1}.  We see that there is a mild decrease in $f_{\pi}^0$ as $\mu$ increases.  
For large values of $\mu$ we see that $d$ must be increased dramatically in order to generate the same degree of dynamical 
symmetry breaking given by $<\ov{q}q>^0$.  Conversely, for small values of $\mu$ ($<0.7\mathrm{GeV}^2$), $d$ must be smaller 
than the quenched value. This is not unnatural if one considers that in the model the typical scale is $\sim1\mathrm{GeV}$ 
-- renormalisation at a scale greater larger or smaller is not physical.  From the lattice results displayed in 
Fig.~\ref{fig:YM-dress} and the results of the full-SDE truncation shown also in Fig.~\ref{fig:alpha-delta}, it is 
expected that the unquenching of the system reduces the gluon dressing by $\sim10\%$ and this indicates that 
$\mu\sim1.0\mathrm{GeV}^2$ is a physical choice.
\begin{table}[t]
\caption{\label{tab:par1}Parameter sets and calculated $<\ov{q}q>^0$ and $f_{\pi}^0$ (chiral limit) using the 
phenomenological model.  $\w=0.5\mathrm{GeV}$ is held fixed.}
\begin{center}\begin{tabular}{|c|c|c|c|c|}\hline
&$d$ ($\mathrm{GeV}^{-2}$) & $\mu$ ($\mathrm{GeV}^2$) & $-(<\ov{q}q>^0)^{1/3}$ ($\mathrm{MeV}$) & $f_{\pi}^0$ 
($\mathrm{MeV}$)\\\hline
quenched&16.0&---&251.2&118.9\\\hline
          &13.0&0.2&251.4&119.2\\
          &15.0&0.5&251.8&118.9\\
unquenched&16.2&0.7&251.4&118.3\\
($N_f=3$) &18.0&1.0&251.2&117.7\\
          &24.0&2.0&251.0&116.2\\
          &45.5&5.0&250.9&113.1\\\hline
\end{tabular}\end{center}\end{table}

We present parameter sets and meson mass results for the physical ($N_f=2+1$) and degenerate ($N_f=3$) unquenching 
scenarios with varying values of $\mu$ in Table~\ref{tab:par2}.  From the insensitivity of the rho meson mass and 
the kaon leptonic decay constant, it is clear that fitting $\mu$ to either of these results is not a good procedure 
and the preceeding discussion of the interaction strength is more relevant to constrain $\mu$.  The near-invariance 
of the meson masses with respect to $\mu$ is the remnant of the renormalisation scale invariance that observables should 
adhere to and that we observe it here is a good sign that the phenomenological treatment retains the character of the 
full theory, insofar as these observables are concerned.
\begin{table}[t]
\caption{\label{tab:par2}Parameter sets and resultant $m_{\pi}$, $f_{\pi}$, $m_K$ $f_K$ and $m_{\ro}$ for the 
phenomenological model.  $\w=0.5\mathrm{GeV}$ is held fixed.}
\begin{center}\begin{tabular}{|c||c|c|c|c||c|c|c|c||c|}\hline
&$\mu$($\mathrm{GeV}^2$)&$d$($\mathrm{GeV}^{-2}$)&$m_u$($\mathrm{MeV}$)&$m_s$($\mathrm{MeV}$)&$m_{\pi}$($\mathrm{MeV}$)
&$f_{\pi}$($\mathrm{MeV}$)&$m_K$($\mathrm{MeV}$)&$f_K$($\mathrm{MeV}$)&$m_{\ro}$($\mathrm{MeV}$)\\\hline
quenched  &---&14.92&5.265&117.55&139.6&130.7&493.7&152.4&737.6\\\hline
unquenched&1.0&16.66&5.185&116.4 &139.6&130.7&493.7&153.0&735.0\\
($N_f=2+1$)   &0.7&15.12&5.22 &117.0 &139.6&130.7&493.7&152.8&736.2\\\hline
          &1.3&18.67&5.14 &5.14  &139.6&130.7&---  &---  &733.2\\
unquenched&1.0&16.94&5.17 &5.17  &139.6&130.7&---  &---  &734.7\\
($N_f=3$)  &0.7&15.15&5.22 &5.22  &139.6&130.7&---  &---  &735.8\\\hline
\end{tabular}\end{center}\end{table}

\begin{figure}[t]
\vspace{1cm}
\mbox{\epsfig{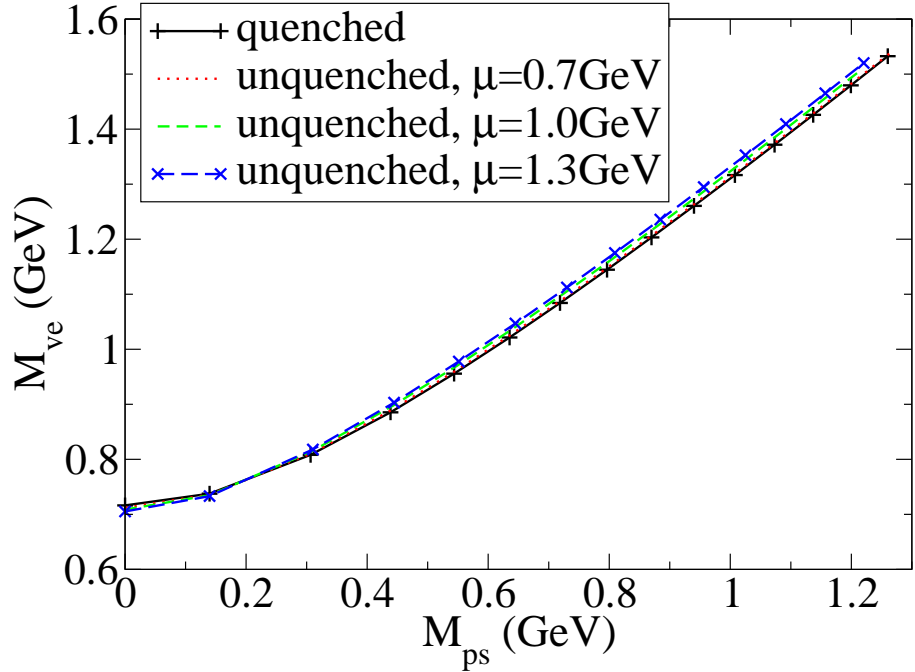}}\\
\caption{\label{fig:mudep}Vector meson masses as a function of pseudoscalar meson masses in the phenomenological model for 
different values of the renormalisation scale $\mu$.  We compare the quenched and degenerate unquenched cases.}
\end{figure}

Finally (and for the sake of completeness), using the quenched and degenerate ($N_f=3$) unquenched quark configuration 
scenarios (using the parameters of Table~\ref{tab:par2}) we plot the vector meson mass as a function of the pseudoscalar 
meson mass in Fig.~\ref{fig:mudep} for increasing quark mass parameter with different values of $\mu$ (cf. Fig.~\ref{fig:plot4}). 
 Between all the cases it is seen that the unquenching is a small, though non-negligible effect.  For $\mu=0.7\mathrm{GeV}^2$, 
 where the parameter $d$ (from Table~\ref{tab:par2}) is almost unchanged from the quenched case the unquenched results are 
 almost identical to their quenched counterpart.  For larger values of $\mu$ the effect due to unquenching becomes larger.



\end{document}